\documentstyle[aaspp4,tighten]{article}

\begin{document}

\title{Population Syntheses for Neutron Star Systems with Intrinsic Kicks}
\author{Chris Fryer, Adam Burrows, and Willy Benz}
\affil{Steward Observatory, University of Arizona, Tucson, AZ 85721}
\authoremail{cfryer@as.arizona.edu}

\begin{abstract}

We use a Monte Carlo binary synthesis code to model the formation and 
evolution of neutron star systems including high-mass X-ray binaries, 
low-mass X-ray binaries, double neutron star systems and radio pulsars.  
Our focus is on the signature imprinted on such systems due to natal  
kicks to neutron stars over and above that imparted by 
orbital motions.  The code incorporates the effect of the galactic 
potential (including rotation) on the velocities of these systems.  
A comparison between 
our models and the observations leads us to infer mean natal kicks 
$\gtrsim 400-500$ km/s.  Moreover, to be consistent with all the 
data, we require a bimodal kick distribution with one peak in the 
distribution near 0 km/s and the other above 600 km/s.  

\end{abstract}

\keywords{stars: neutron; pulsars; supernova; binaries}

\section{Introduction}

Support for the claim that many neutron stars move with high velocities 
continues to mount.  Probably the most compelling evidence lies in the 
transverse velocities of the radio pulsar population.  
With the most recent proper motion 
measurements and the newly-corrected distance determinations of Taylor \& 
Cordes (1993), Lyne \& Lorimer (1994) have derived a mean pulsar velocity 
of $450$ km/s, an increase of almost a factor of 2 over previous 
estimates.  Velocities above $800$ km/s have been inferred 
by associating pulsars with supernova remnants (Caraveo 1993; Frail, Goss 
\& Whiteoak 1994 although see Gaensler \& Johnston 1995) 
and from observations of the bow shocks produced by neutron 
stars as they plow through the interstellar medium (Cordes, Romani, \& 
Lundgren 1993).  Neutron star kicks have been invoked to explain 
characteristics of  O/B runaway stars (Leonard \& Dewey 1992), double neutron 
star (DNS) systems such as PSR B1913+16 and PSR B1534+12 (Flannery 
\& van den Heuvel 1975; Burrows \& Woosley 1986; Yamaoka, Shigeyama, 
\& Nomoto 1993; Fryer \& Kalogera 1997), the non-zero angle between the spin 
and orbit axes of some recycled pulsar systems 
(Kaspi {\it et al.} 1996; Wasserman, 
Cordes, \& Chernoff 1996), galactic gamma-ray bursters 
(Colgate \& Leonard 1994; Podsiadlowski, Rees, \& Ruderman 1995; 
Lamb 1995), and highly eccentric Be/NS binaries 
(van den Heuvel \& Rappaport 1986).

Distilling the actual kick imparted to neutron stars from the 
pulsar velocities requires the delicate removal of the forces that have 
modified the pulsar motion since birth.  For example, if the progenitor 
of the pulsar was in a binary system prior to the supernova 
explosion which formed the pulsar, then the binary rotational 
velocity would affect the final pulsar motion.  In addition, escape from 
the binary potential, the effects of galactic rotation and the motion 
through the galactic potential all modify the pulsar velocities.  
The most reliable approach to this distillation process is the simulation 
of the pulsar velocity distribution given a kick distribution and a model for 
the binary and galactic effects.  Using Monte Carlo binary population 
synthesis simulations, Dewey \& Cordes (1987) found that $100-150$ km/s 
kick velocities were required to explain the pulsar transverse-velocity 
distribution with the old distance measurements.  More recently, 
Iben \& Tutukov (1996) have claimed that this old pulsar transverse-velocity 
distribution can be explained by binary effects alone.  Iben \& Tutukov 
use much more detailed binary population synthesis code which includes 
the synthesis of low-mass X-ray binaries (LMXBs) and high-mass X-ray binaries 
(HMXBs) (Iben, Tutukov, \& Yungelson 1996a, 1996b) but which does not include 
the effects of neutron star kicks.  To explain the transverse velocity 
distribution of pulsars using the {\it latest} proper motion measurements 
and the {\it newest} distance corrections, our calculations which include 
binary and galactic effects require large neutron-star kicks ($ > 400$ km/s).  

The radio pulsar velocity data alone provide evidence for kicks, 
but they are unable to restrict the kick distribution itself.  
We show that Maxwellian distributions, flat distributions, even a delta 
function kick distribution, will match the observed radio pulsar data.  
To learn more about the kick distribution, we follow the approach of 
Iben \& Tutukov (1996) and expand our study to include additional 
neutron star populations:  low-mass X-ray binaries (LMXBs), high-mass 
X-ray binaries (HMXBs), DNSs, and globular cluster neutron stars.  
Naturally, any kick distribution is also restricted in that it must 
produce roughly the correct numbers of each of these neutron star 
systems.  High kicks will overcome the gravitational potential of 
binaries and globular clusters preventing the formation of LMXBs, 
HMXBs, DNSs and globular cluster neutron stars.  Our basic approach 
is to calculate the minimum birthrate of each type of these systems based 
on the observations and the maximum birthrate derived from our Monte 
Carlo code for a given kick distribution which fits the pulsar 
velocity distribution.  If this maximum birthrate is less 
than the observed minimum, we can ``conservatively'' 
conclude that the input kick distribution, although it fits the pulsar 
velocity distribution, is not the correct kick distribution.  
Employing the constraints of all these 
systems, we can rule out many of the kick distributions in the literature
(Kalogera \& Webbink 1996a - hereafter KW96; Brandt \& Podsiadlowski 1995).
We do find that a two-population pulsar velocity distribution is able 
to fit the current observations and our best-fitting double-peaked distribution 
has one peak near $\sim 625$ km/s and the other near $0$ km/s\footnote{
The peak near 0 km/s may be significantly broadened and still satisfy 
our constraints.  A flattened distribution ranging from 0 km/s - 200 km/s 
will fit the data equally well.}.  
This bimodal kick distribution is a challenge to any theory of kick 
origins.  

In \S 2, we discuss the observed pulsar distribution and our method 
to limit the effects of biases.  We then outline our technique to 
compare our simulations to the data.  We repeat this process for 
the neutron stars in globular clusters and the massive binary 
systems.  In \S 3, we describe the various uncertainties in our population 
synthesis calculations and show the dependence of our results 
on the many poorly constrained binary population synthesis parameters.  
In \S 4, we discuss the results and implications of these 
new simulations and the constraints they place on the neutron star kick 
velocity distribution.  We conclude with a discussion of our results.    

\section{Neutron Star Populations}

We determine possible nascent neutron star kick distributions by 
combining the radio pulsar proper motion data with the retention 
fraction of neutron stars in globular clusters and the formation 
rates of LMXBs, HMXBs, DNSs.  We begin by discussing the birthrate 
comparisons for the massive binary systems and conclude this section 
discussing the globular retention fraction, pulsar velocity 
distribution, and lastly, the formation mechanisms of O/B runaway 
stars. 

For the binary populations, we must 
carefully determine the birthrate (BR) of each type of binary system, 
given by:
\begin{equation} \label{eq:br}
BR = \frac{{\rm Number \; of \; Systems}}{SN} \times \frac{SN}{yr} \times f_{binary}.  
\end{equation}
Using our Monte Carlo code, we simulate $10^6$ binary systems and calculate the 
number of each type of system produced and the total number of supernovae 
($SN$), from which we derive the first term in equation (1).  By assuming a 
supernova rate ($SN/yr$) and the fraction of systems in binaries, $f_{binary}$, 
we calculate the birthrate of each type of system.  

The birthrate itself can not be easily compared with the 
observations.  Hence, we resort to a variety of indirect techniques to 
constrain the models.  Since we use a different technique for each type of system, 
we discuss each of them separately.  When we encounter any uncertainty in 
a calculation, we choose the conservative bound such that our simulated birthrates 
are always upper limits.  

There are a number of ways to estimate formation rates of massive systems, the most 
direct being the supernova rate.  To compare our rates with the recent 
population synthesis work, we list the supernova rates of previous population 
synthesis work given the assumptions 
in their calculations:  KW96 - $\sim 7.7 \times 10^{-3} \, {\rm SN \, yr}^{-1}$, 
Dalton \& Sarazin (1995 - DS95) - $\sim 7 \times 10^{-3} \, {\rm SN \, yr}^{-1}$, 
and Iben, Tutukov \& Yungelson (1996a, 1996b) - $\sim 9.1 \times 10^{-3} \, 
{\rm SN \, yr}^{-1}$.  However, galactic SN rates 
themselves are typically estimated to be in the range 
$10^{-2}-3 \times 10^{-2} \, {\rm SN \, yr}^{-1}$ (Tammann, Loeffler, \& Schroeder 1994).  
To be consistent the results of Tammann {\it et al.} (1994), while choosing 
a value close to the recent binary population synthesis work, 
we assume a supernova rate of $10^{-2} \, {\rm SN \, yr}^{-1}$ 
for our simulations.  The binary fraction ($f_{binary}$) depends upon the mass 
ratio distribution of binaries that we adopt.  We discuss the 
various mass ratio distributions and their resultant binary fractions in \S 3.

\subsection{Low-Mass X-ray Binaries}

The X-ray emission of an LMXB is powered by Roche-lobe overflow 
from its low-mass companion (Iben, Tutukov, \& Yungelson 1996b).  Although a necessary 
condition for LMXB creation is that the low-mass companion remain bound to  
the neutron star after the supernova explosion, this condition is not 
sufficient.  The bound system must evolve to a phase in which stable 
Roche-lobe overflow occurs.  For our simulations, we use the technique outlined by 
KW96.  We evolve the orbital separation by both gravitational radiation and 
magnetic braking.  Using the results of Kalogera \& Webbink (1996b), we 
restrict our sample to those systems that develop stable sub-Eddington 
or super-Eddington accretion.  Since systems with super-Eddington accretion 
may not be observed as LMXBs and since systems with super-Eddington accretion rates 
are an order-of-magnitude more numerous than sub-Eddington accretion, the inclusion 
of super-Eddington systems may overestimate the birthrate of LMXBs 
by over a factor of 10.  

The astute reader may worry that our simulations do not consider 
all of the possible formation scenarios for LMXBs and, hence, that we 
are underestimating their formation rate.  Our simulations include the standard 
formation scenario which uses a common-envelope phase to tighten the pre-SN 
orbit (van den Heuvel 1983).  In addition, we include scenarios which 
use the kick to reduce the orbital separation after the supernova 
without the aid of a common-envelope phase (Kalogera 1996).  We do 
not include any scenarios involving a Thorne-$\dot{\rm Z}$ytkow phase (Eggleton 
\& Verbunt 1986).  Models of common-envelope evolution (Chevalier 1993, 
Brown 1995, Fryer, Benz, \& Herant 1996) show that the neutron star 
would collapse into a black hole before it could spiral into the core 
of its companion to form a Thorne-$\dot{\rm Z}$ytkow object.  Similarly, we ignore 
the accretion-induced collapse (AIC) route to LMXBs.  
Simulations by Woosley \& Baron (1992) limit the total rate of AICs to 
$10^{-4}$ yr$^{-1}$ to avoid nucleosynthetic contamination by their ejecta.  
This rate is comparable to the rate predicted by Iben, Tutukov, \& Yungelson 
(1996b), who conclude that AICs make up no more than a few percent of the LMXB 
population.  In addition, recent simulations by Fryer {\it et al.} 
(1997) suggest an upper limit on the total AIC rate an order of magnitude 
lower than that of Woosley \& Baron.  
Hence, it is unlikely that AICs contribute to the LMXB population.  

Using our Monte Carlo calculations, we determine the number of LMXBs formed per 
supernova (the first term in equation 1).  To compare 
with the observations, we would like to multiply our birthrate 
with the lifetime of our LMXBs to derive the number of galactic LMXBs.  
However, the lifetime of an LMXB is very difficult to estimate.  We instead  
calculate an upper limit to the X-ray flux of these systems collectively 
by assuming that the low-mass companion 
is completely consumed by the neutron star and that all of the energy from mass 
accretion is converted into X-ray photons.  This galactic luminosity is:  
\begin{equation}
L_{X}^{tot}=\frac{G M_{NS}}{R_{NS}} \times \sum_{M_{co}}\,BR (M_{co})\,
\times M_{co}, 
\end{equation}
where $G$ is the gravitational constant, $M_{co}$ is the mass of the 
companion which is accreting onto the neutron star, and $M_{NS}$ and $R_{NS}$ are the 
mass and the radius of the neutron star, respectively.  $BR$ is the LMXB 
birthrate calculated in our Monte Carlo simulations (eq. 1).  KW96 estimate the X-ray 
flux from galactic LMXBs to be $L_{X,tot} \sim 1.7 \times 10^{39}$ 
erg s$^{-1}$.  In \S 4, we make use 
of the fact that we have overestimated both the LMXB formation rate and 
the LMXB X-ray emission.  If, for a given kick distribution, our calculated upper 
limit falls below the observed value, that kick distribution is excluded.  

\subsection{High-Mass X-ray Binaries}

HMXBs are thought to be powered by material from the massive companion's 
wind.  For these systems, we use the same technique as DS95.  Using the 
models of Schaller {\it et al.} (1992), we determine the high-mass companion's 
radius, its mass loss rate, and wind velocity as a function of the 
companion's age, from which we can estimate the mass accretion rate 
onto the neutron star as a function of time (see DS95).  Then 
the maximum X-ray luminosity of each HMXB can be estimated:  
\begin{equation}
L_{X}=\frac{G \dot{M}_{acc} M_{NS}}{R_{NS}}
\end{equation}
where $\dot{M}_{acc}$ is determined in the same manner as in 
DS95.  DS95 introduce an efficiency parameter for the conversion 
of potential energy into X-ray photons, but we assume, as we 
do in the case of LMXBs, that the conversion is $100$\% when 
calculating an upper limit to the formation rate of HMXBs.  

Using the Schaller {\it et al.} (1992) models, we not only calculate the X-ray 
luminosity, but the HMXB lifetime.  For 
HMXBs, we can combine our estimated birthrate with this lifetime to 
determine a total HMXB population.  We compare our brightest 
sources with the bright galactic sources.  Meurs \& van den Heuvel 
(1989) estimate the number of HMXBs with $L_{X} > 10^{36} \, {\rm erg \, s}^{-1}$ 
to be $55 \pm 27$.  We will require the upper limit from our simulations 
to give $N(L_X>10^{36} {\rm erg s^{-1}}) \gtrsim 28$.  As in the case for LMXBs, if our 
simulated upper limit falls below the lower limit in the observations, 
we conclude that the kick distribution does not fit the data.  

\subsection{Double Neutron Star Systems}

We calculate the birthrate of DNSs by including all bound double 
neutron star systems.  This simple prescription is suggested by the 
fact that there are many opportunities during the evolution of these 
systems to tighten their orbits and recycle their pulsars.  
van den Heuvel \& Lorimer (1996) estimate that the formation 
rate for DNS systems whose merger timescale is less than a Hubble time
is $\sim 8 \times 10^{-6} {\rm y}^{-1}$, but previous estimates 
were as low as $1-3\times 10^{-6} {\rm y}^{-1}$ (van den Heuvel 1995, 
Curran \& Lorimer 1995).  Since we are interested 
in estimating an upper limit on the theoretical birth rate, we require only 
that our total number of bound DNS systems be greater than 
the $\sim 10^{-6} {\rm y}^{-1}$ predicted by the observed close binary systems.  

Alternate scenarios for double neutron star formation do exist.  In globular 
clusters, a viable formation scenario for DNSs such as PSR 2127+11C involves 
stellar collisions with binaries in cluster cores (Anderson {\it et al.} 1990).  
This scenario is not 
a likely formation mechanism for the galactic disk DNSs.  Brown (1995) 
has suggested an alternate formation scenario which includes a 
double helium star phase.  This mechanism requires that the binary 
components have nearly equal masses.  Our code models these 
systems, but for the mass-ratio distributions we use (see \S 3), 
this mechanism provides a negligible contribution to our DNS formation 
rate.

\subsection{Globular Clusters}

Rather than calculate a birthrate of neutron stars in globular clusters, 
we simply estimate their retention fraction for a given kick velocity.  
The large population of neutron stars observed in globular clusters requires that a 
sizable fraction ($>1-10$\%) of their neutron stars remain bound to the cluster 
(Bhattacharya \& van den Heuvel 1991).  If they form primarily from 
core-collapse supernova in situ, then the retention fraction places 
useful constraints upon the neutron-star 
kick distribution.  An alternate formation mechanism in globular 
clusters involves the accretion-induced collapse of white dwarfs 
(e.g. Bailyn \& Grindlay 1990).  The role AICs play in globular 
clusters is restricted, just as with LMXBs, by the nucleosynthetic 
yields predicted by Woosley \& Baron (1992) and Fryer {\it et al.} (1997).  
Bailyn \& Grindlay (1990) require an AIC rate close to $10^{-4}$ y$^{-1}$ 
to explain the neutron stars in globular clusters, an order of magnitude higher 
than the upper limit given by Fryer {\it et al.} (1997).  It is therefore unlikely 
that AICs make up more than about 10\% of the neutron stars in globular 
clusters.  

However, if neutron stars created through core-collapse explosions do indeed 
receive large kicks, Drukier (1995) has shown that the 
retention fraction of these neutron stars can be quite low ($\lesssim 1 \%$ 
of the neutron stars formed).  
We use the retention fractions versus neutron-star velocity derived 
by Drukier (1995) to determine the retention fractions 
of neutron stars (both bound and unbound systems) for all of our kick 
distributions.  Drukier uses both Fokker-Planck and Mitchie-King models 
to simulate the range of globular cluster retention fractions and includes 
specific models for NGC 6397 and $\omega$ Cen.  Our limit for a satisfactory 
neutron-star kick distribution requires that 1\% of the neutron stars formed 
(both binary systems and single stars are considered) 
in NGC 6397 remain bound.  We calculate the entire range of retention fractions 
derived using the Drukier models.  

\subsection{Radio Pulsars}

The primary constraint on the neutron star kick distribution are 
the pulsar transverse velocities.  The high velocities, as we show in 
\S 4, require high-velocity kicks.  Given the strong dependence 
of our results upon the proper motion velocities, we must first 
discuss the many uncertainties surrounding any interpretation of this 
data.  We conclude this section with a discussion of our simulations.

\subsubsection{Observations and Uncertainties}

The current database of pulsars with proper motions now contains 
well over 100 pulsars (Taylor, Manchester, \& Lyne 1993).  However, 
this sample contains many biases and uncertainties (Iben \& Tutukov 
1996; Hansen \& Phinney 1997; Cordes \& Chernoff 1997).  In this 
section, we discuss many of the uncertainties and biases listed 
in the literature and describe our method to limit their effects.  

Our pulsar sample is taken from the proper motion data of Taylor, 
Manchester, \& Lyne (1993) with distances determined using the new 
electron density model of Taylor \& Cordes (1993).  For some pulsars, 
the distances estimated from the new electron density model are over a 
factor of two greater than their old predictions.  
Alternate distance estimation techniques provide some support for 
the Taylor \& Cordes distances, but there exist specific cases, such as 
PSR J0738-4042, where the preferred distance is 5 times smaller than that 
predicted by the electron density model (Johnston {\it et al.} 1996).  The 
distance estimated by Taylor \& Cordes for PSR J0738-4042 
was quite high ($>11$ kpc).  By restricting our sample to the radio pulsars 
within $5$ kpc of the sun, we hope to avoid the most grevious distance errors.  
Nonetheless, distance measurement errors are a major 
concern and we will discuss their effect on our results in \S 4.  

Aside from uncertainties in the distance estimates, we must be 
careful to avoid any biases in our radio pulsar sample.  A clear selection 
bias is that very high-velocity pulsars rapidly leave the galactic 
disk and can even escape the galactic potential.  Low-velocity pulsars, on the other hand, 
remain bound to the disk and are easily detected.  We avoid this bias 
by limiting our sample to the young radio pulsar population  
($t_{age} = P/ 2 \dot{P} < 3\times 10^{6}$ y).  In addition, by restricting our 
sample to those pulsars whose ages are less than the typical 
luminosity decay times (Gunn \& Ostriker 1970), we avoid uncertainties in 
the pulsar age-luminosity relation.  

Iben \& Tutukov (1996) have suggested that there may be a bias against 
low velocity pulsars.  The proper motion of a distant, low-velocity pulsar is 
difficult to determine.  Iben \& Tutukov (1996) have claimed that there 
is a trend in the data supporting this hypothesis (see Figure 1).  
This bias is not real, but instead is probably due to errors in the 
distance estimates and is akin to the luminosity/velocity relation suggested by 
Tutukov, Chugai, \& Yungelson (1984) and Hansen \& Phinney (1997).  They noted a clear 
trend in the data showing that the lowest velocity pulsars have 
lower luminosities.   By realizing that $v_{trans} \propto$ proper motion 
$\times$ distance and that Luminosity $\propto$ distance$^{2}$, Dewey \& Cordes (1987) 
argued that this trend was not a bias, but a consequence of distance errors.  
This effect is illustrated in Figure 2.  Distance errors also explain the trend upon 
which Iben \& Tutukov (1996) base their selection bias (see Fig. 3).  
By misinterpreting these trends as true biases rather than straightforward 
distance errors, Tutukov, Chugai, \& Yungelson (1984), Iben \& Tutukov 
(1996), and Hansen \& Phinney (1997) predict much lower mean pulsar 
velocities ($\sim 100-300$ 
km/s) than those predicted by Lyne \& Lorimer (1994).  Given that the 
Dewey \& Cordes (1987) argument fits the observations so well (Figs. 2,3), 
the bias is not likely to be true, and hence, like Lyne \& Lorimer (1994), 
we do not correct for it.\footnote{However, the current sample of pulsar 
proper motions is far from complete and many of the issues that 
Hansen \& Phinney (1997) bring up may prove to be important.}

Thus, we limit our data both in distance ($<5$kpc) and age ($t_{age} 
< 3\times 10^{6}$ y), taking only those pulsars which satisfy both 
of these constraints.  Unfortunately, this limits our sample of pulsars 
to 27 pulsars with proper motion estimates.  Figure 4 shows the distribution 
of transverse velocities for this sample where we have smoothed the data 
by assuming a distance error of 30\%.

\subsubsection{Models}

Our results rely upon several assumptions about the kick distribution:  that 
it is isotropic and it does not depend upon the binary nature of the exploding 
star (i.e. The kick distribution for neutron stars formed in binary systems 
is identical to the distribution for neutron stars formed in supernova explosions 
of single stars.).
Using Monte Carlo statistics, we compare our simulated pulsar population\footnote{
unbound neutron stars produced in our synthesis calculations either 
because their progenitor was a single star or because they became unbound 
during the supernova explosion in the binary system}
to the radio pulsar sample, constrained by our age ($t_{age}< 3\times10^6$ y) 
and distance ($D_{sun}<5$ kpc) limits.  For each kick distribution, we calculate 
a pulsar velocity distribution with our population synthesis code which includes 
both binary systems and single stars.  To calculate the transverse velocity 
distribution to compare with the observations, we must follow the motions 
of the pulsars in a galactic potential.  

We use the galactic potential of Miyamoto and Nagai (1975) 
\begin{equation}
\Phi(R,z)=\frac{G M_{gal}}{({R^{2}+[a+(z^2+b^2)^{1/2}]^{2}})^{1/2}}
\end{equation}
where $R$ is the distance from the galactic center in the plane of the 
disk, $M_{gal}$ is the mass of the galaxy, 
$z$ is the distance off the disk and $G$ is the gravitational constant.  
We use fits by Miyamoto and Nagai for $a$ and $b$ ($a=7.258 $kpc and 
$b=0.520 $kpc).  We normalize $M_{gal}$ by insuring that the rotational 
velocity of the sun at $8.5$ kpc is $225$ km/s.  

We distribute our initial binary systems randomly following the O/B disk population 
(Mihalas \& Binney 1968) with a disk scale length of $3.5$ kpc and  
a scale height and cutoff out of galactic plane of $60$ pc and $300$ pc, respectively.  
The motion of each system consists of a component from the galactic rotation 
and a randomly oriented velocity due to binary and kick effects.  Figure 5 
shows the effects, first of binary evolution, and then the effects of the 
galactic potential including rotation, upon a kick distribution.  The effect 
of the galactic potential not only leads to significant changes in the mean 
pulsar velocity (for some kick distributions, the deviation can be as high 
as 50\%), but it drastically alters the pulsar velocity distribution and 
can not be ignored.  

We use the Kolmogorov-Smirnov test to derive the 
probability that the simulated pulsar velocity distribution, subject to 
the same age and distance constraints as our observed sample, is not from the 
same parent population as the radio-pulsar sample.  We repeat the test 
using the extreme possible velocities for each pulsar (by including 
both distance and proper motion errors\footnote{Note that nearly 
20\% of the pulsars have proper motion errors greater than the 
measured value.  That is, their proper motions are, within the 
errors, zero. (Taylor, Manchester, \& Lyne 1993)}).  
As a secure limit, we exclude only those distributions where 
this probability is greater than 99\%.

\subsection{O/B runaway stars}

O/B runaway stars are O or B stars that have been given high space velocities.  
One proposed mechanism for O/B runaway stars is ejection 
during a supernova event (Blaauw 1961).  However, this mechanism requires 
that the O/B runaway stars remain bound to the newly-formed neutron star 
(Leonard 1990, Leonard \& Dewey 1992) and the current observational evidence 
suggests that most O/B stars are not in close binaries (Gies \& Bolton 1986, 
Sayer, Nice \& Kaspi 1996, Philp {\it et al.} 1996).  An alternate formation 
mechanism for these objects is dynamical ejection in cluster environments 
(Leonard 1995) and this mechanism may well explain most O/B runaway stars.  
Because no standard model exists for the formation of these objects, we 
do not use them as a constraint for our kick distributions.  In the conclusions, 
we briefly address these objects in the context of our derived kick distribution. 

\section{Simulations}

To determine the birthrate of each type of neutron star system, we first 
calculate the number of systems formed per supernova, the first 
term in equation (\ref{eq:br}).  To calculate this term, we have created 
a Monte Carlo population synthesis code which chooses from a range of 
initial conditions and then evolves the binary system through one, 
and if the secondary mass is sufficiently high, a second supernova explosion.  
A variety of uncertainties and ``free-parameters'' (both 
in the initial conditions and in the subsequent binary evolution) results in a 
broad range of birthrates.  Therefore, to attack the problem of 
neutron star kicks, we must explore the realistic range in these rates.  
In this section, we present the results of an intensive study 
of the effects of the initial conditions and free parameters on 
the production rates of LMXBs, HMXBs and DNSs, and on the radio pulsar velocity 
distribution.  The results for the different populations are summarized in Figure 6.    
Although the birthrates for these systems can change by over an order of magnitude 
as we vary the parameters, if the kick is sufficiently strong, it will dominate 
the pulsar velocity distribution (Figure 7).  The mean pulsar velocity, 
using a delta-function kick velocity of 200 km/s,  
after binary effects ranges from $199$ km/s to $202$ 
km/s.  The 
velocity dispersion ranges from $11-15$ km/s 
for all the binary parameters 
except if we vary the common envelope effeciency or mass loss.  For a common 
envelope efficiency $\alpha=0.2$, the dispersion is $5$ km/s and for no mass 
loss, the dispersion is $22$ km/s.  The variation in the velocity dispersions 
is so small that, for the purposes of our simulations, any set of binary parameters 
will essentially give the same pulsar velocity distribution.  (This is not the 
case when the kick magnitude becomes smaller than the average orbital velocity 
$\lesssim 50$ km/s.)

\subsection{Initial Conditions}

Four parameters are required to describe a binary system.  These 
are the masses of the two stars, $M_{p,0}$ and 
$M_{c,0}$, the orbital separation, $A_{0}$, and the initial 
eccentricity, $e_{0}$.  Unfortunately, for massive binaries, 
observational data only moderately constrain these 
parameters (Hogeveen 1991).  Therefore, we are forced to consider a wide range of initial 
conditions and to use the neutron star binary production rates themselves 
to limit the initial conditions.   The varying effects of these 
assumptions are summarized in Table 1, which lists the birthrates 
of LMXBs, HMXBs, and DNSs for 3 delta-function kick velocities 
and a standard set of parameters.  For each kick 
velocity, it also shows the effects of a variety of deviations 
from the standard parameter set.

\subsubsection{Mass Ratios and the Initial Mass Function}

In our simulations, we determine the initial mass of the primary ($M_{p,0}$) 
by sampling an 
Initial Mass Function (IMF):
\begin{equation}
f(M_{p,0}) \propto M_{p,0}^{-\alpha_{IMF}}.
\end{equation}
We retain $\alpha_{IMF}$ as a free parameter, but must choose a minimum 
and a maximum neutron star forming primary mass.  For most of our simulations, we  
use $\alpha_{IMF}=2.7$ (Scalo 1986) and primary mass limits of $10$ and 
$40 M_{\odot}$.  

The companion mass distribution is much more difficult to determine.  The 
standard technique prescribes a mass ratio ($q=\frac{M_{s}}{M_{p}}$) distribution 
$P(q)$ by 
\begin{equation}
P(q) \propto q^{-\alpha_{MR}}.
\end{equation}
Observational data for massive star binaries is limited and the effects 
of selection biases can be extreme.  Garmany, Conti, and Massey (1980) claim 
a strong, bias-corrected, peak at $q=1$.  This led DS95 to choose 
$\alpha_{MR}=-1$ for the bulk of their simulations.  However, by accurately accounting 
for the selection biases, Hogeveen (1991) found that the Garmany {\it et al.} results 
vastly underestimate the number of low-mass companions.  Hogeveen favors a mass ratio 
distribution which is peaked at low $q$ values with $\alpha_{MR}=2.7$ which flattens 
to $\alpha_{MR}=0$ below some critical $q=q_{0}$.  
We use a range of values for $\alpha_{MR}$ and $q_{0}$.  In Table 1, we see 
that low values of $\alpha_{MR}$ such as those given by Garmany, Conti, and 
Massey (1980) lead to a maximum in the DNS production rate.  However, the higher 
value of $\alpha_{MR}$ claimed by Hogeveen (1991) is required to explain the 
production rate of LMXBs.  For most of our simulations we use the  
high $\alpha_{MR}=2.7$ value, and vary only the critical value $q_0$. 

The binary fraction depends upon the choice for the mass ratio distribution 
parameters.  For $\alpha_{MR}=2.7$ and $q_{0}=0.35$, Hogeveen (1991) gives a 
binary fraction of 35\%.  For $\alpha_{MR}=2.7$ and $q_{0}=0.15$, this value 
increases to $\sim 65$\%.  For the mass ratio distributions derived by Garmany, 
Conti, \& Massey (1980), we use their calculated binary fraction of 43\%.  

\subsubsection{Orbital Parameters}
The distribution of initial eccentricities and separations is also not well 
known for massive systems.  For then initial orbital separation ($A_{0}$), 
we assume with Kraicheva {\it et al.} 
(1979) that
\begin{equation}
P(A_{0}) \propto 1/A_{0}.
\end{equation}
We use an inner separation of twice the initial primary radius and a range of 
outer separations ($10^{4-6} R_{\odot}$).   For initial eccentricity 
($e_0$), we choose 
two distributions:
\begin{equation}\label{eq:pdel}
P(e_{0})=\delta(e_{0})
\end{equation}
and
\begin{equation}
P(e_{0})=1.
\end{equation}
For most of our simulations, we use an outer separation of $10^{4}  R_{\odot}$ 
and the eccentricity distribution of eq. (\ref{eq:pdel}).  As can be seen 
in Table 1, the choice of these has very little effect upon the 
neutron star system production rates.  

\subsection{Stellar Models and Binary Evolution}
We base our binary evolution calculations on stellar models of single stars,  
to which we add binary effects such as mass transfer and common envelope 
evolution.  For 
stellar radii and masses at different evolutionary periods, we use 
the fits from KW96 of the massive stellar models of Schaller {\it et al.} 
(1992) and the helium star models of Habets (1985) and Woosley, 
Langer, \& Weaver (1995).  Although many aspects of binary evolution 
are not well understood, the uncertainties have, either rightly or wrongly, 
been lumped into a few categories.  Chief among these 
are mass transfer, common envelope evolution, and stellar winds.  
The varying effects of these assumptions are summarized in Table 2, which 
lists the birthrates of LMXBs, HMXBs, and DNSs for 3 delta-function kick velocities 
and a standard set of parameters.  For each kick 
velocity, it also shows the effects of a variety of deviations 
from the standard parameter set.

When the primary star overfills its Roche Lobe, mass transfer begins.  
For binary systems with mass ratio $q<0.4$, we assume that there is no stable 
mass transfer (Webbink 1979; Yungelson \& Tutukov 1991; van den 
Heuvel 1983) and that the system immediately goes into a common envelope.  
For systems with less extreme mass ratios, we assume, 
as did DS95, that the mass transfer is initially stable.  When the two 
stars attain equal masses, it is assumed that the mass transfer is no longer 
stable and a common envelope phase begins.   

For stable mass transfer, we follow the prescription of van den Heuvel (1995):
\begin{equation}
\Delta M_{s} = - \Delta M_{p} \times (1-f_{trans}) 
\end{equation}
where $\Delta M_{s}$, $\Delta M_{p}$ are the change in mass of the secondary 
and primary star, respectively, and $f_{trans}$ is the fraction of mass lost from 
the primary which does not accrete onto the secondary and is removed from the system.  
From Table 2, we see that the results depend only slightly on $f_{trans}$ and 
for most of the simulations, we use $f_{trans}=0.5$.  
During this phase, the loss of orbital angular momentum is determined by the 
parameterization of de Loore \& De Greve (1992):
\begin{equation}
\frac{\Delta J_{orb}}{J_{orb}}=1-\left ( 1-\frac{\Delta M_{tot}}{M_{tot}} 
\right ) ^{\gamma} ,
\end{equation}
where $J_{orb}$ and $M_{tot}$ are the pre-overflow values.  The value 
of the parameter $\gamma$ is poorly constrained.  
However, as seen in Table 2, uncertainty in $\gamma$ has very little effect 
on the results.  For most of the simulations, we use $\gamma = 2.1$ as 
estimated by De Greve {\it et al.} (1985).  

For common envelope evolution, we assume that no mass is gained by the secondary 
star and that the primary loses its hydrogen envelope.  For the ratio of 
post-common envelope to pre-common envelope binary separation, we use Webbink (1984):
\begin{equation}
\frac{A_{f}}{A_{i}}=\frac{\alpha_{CE}\, r_{L}\, q}{2} \left(\frac{M_{He}}
{(M_{p}-M_{He}) + \frac{1}{2}\alpha_{CE} r_{L} M_{s}} \right),
\end{equation}
where $r_{L}=R_L/A_i$ is the dimensionless Roche lobe radius of the primary (Eggleton 1983),
\begin{equation}
r_{L}=\frac{0.49 q^{-2/3}}{0.6 q^{-2/3} + ln(1+q^{-1/3})}.
\end{equation}
$M_{He}$ is the mass of the primary's helium core and $\alpha_{CE}$ 
represents the efficiency with which orbital energy is injected 
into the common envelope.  The fate of close 
binary systems depends strongly upon this parameter and the current set 
of hydrodynamical simulations (Taam \& Bodenheimer 1991; 
Yorke, Taam \& Bodenheimer 1995; Rasio \& Livio 1996) do not yet provide 
a definitive value for this parameter.  Indeed, $\alpha_{CE}$ is probably a function of binary 
system.\footnote{Iben, Tutukov, \& Yungelson (1996a) use a slightly different 
equation for the post-common to pre-common ratio in which the efficiency 
parameter $\alpha_{CE}$ has a different meaning.}  However, 
as can be seen in Table 2, by choosing a higher efficiency, we increase 
the numbers for all the binary populations.  In our comparison with 
observation, we use our simulations only to provide upper limits and, 
to be conservative, we maximize the numbers by choosing a high efficiency 
($\alpha_{CE}=1$).  

Although the parameter $\alpha_{CE}$ has not yet been constrained by 
hydrodynamical simulations, conclusions can be drawn about specific 
aspects of common-envelope evolution.  For example, Taam \& Bodenheimer 
(1991) found that because He-star giants do not have the steep density 
profiles of the hydrogen counterparts, the likely outcome of a common-envelope 
phase with a He-star giant single star system in which the two stars have 
merged.  In our code, these systems can no longer produce any of the massive 
X-ray binaries, but we retain the object for the pulsar velocities.  However, 
as they make up less than $0.1$\% of the total number of binary systems, 
they have little effect on the pulsar velocity distribution.

The high density medium that surrounds 
neutron star in the common envelope phase cause neutrinos, rather than 
photons, to be the dominant coolant.  Hence, the accretion rate onto 
the neutron star is not limited by the Eddington rate (Chevalier 1993,1996; 
Brown 1995; Fryer {\it et al.} 1996).  For hydrogen giants, angular momentum 
(Chevalier 1996) or explosions induced by neutrino heating (Fryer {\it et al.} 
1996) may restrict the accretion and allow the neutron star to survive 
this phase.  However, in the denser environments of helium star giants, 
angular momentum and neutrino heating will not be sufficient to prevent 
black hole formation.  In our simulations, we assume that neutron 
stars survive hydrogen-giant common envelope phases, but collapse to black 
holes if they progress through a helium-giant common envelope phases.

Mass loss due to stellar winds has a direct effect on stellar mass, 
which, in turn, has a strong effect upon the stellar radius.  For both 
the models of Schaller {\it et al.} (1992) 
and those of Woosley, Langer \& Weaver (1995), we 
parameterize the wind mass loss with 
\begin{equation}
\Delta M_{wind} = f_{wind} \times \Delta M^{models}_{wind}, 
\end{equation}
where $\Delta M^{models}_{wind}$ is the mass loss from the primary through its wind 
as predicted by Schaller {\it et al.} (1992), which agrees reasonably well with 
Woosley {\it et al.} (1995).  The Schaller {\it et al.} mass-loss rates reflect 
upper limits to the mass-loss 
from winds and in our simulations, we consider the range  
$0.0<f_{wind}<1.0$ (see Table 2).  For most of our simulations, $f_{wind}$ is 
set equal to unity.   

\subsection{Pulsar Velocities and Globular Clusters}

An additional set of parameters can be derived if we include 
uncertainties in the galactic and globular cluster potential 
models.  Distributing the pulsar formation position along 
spiral arms rather than a smooth disk may also have an effect 
on the simulated pulsar transverse-velocity distribution.  For 
the purposes of this paper, we do not consider these effects.  
However, we do consider a range of globular cluster potentials 
from Drukier (1995) which lead to a range in retention fraction.  
Since the range in retention fraction depends sensitively upon 
the kick distribution, we present these ranges separately for 
each kick distribution in \S 4.  

\section{Natal Kick Distributions}

Given a kick distribution, we can use our Monte Carlo code to 
derive the production rate of LMXBs, HMXBs, and DNSs as well as 
the pulsar velocity distribution and the globular cluster retention fraction.  
We stress that for all of the neutron-star populations, we 
overestimate the production rate.  The ratio of our simulated rates 
to the actual rates may well be greater than ten (see \S 2).  Similarly, 
our globular cluster retention fractions are upper limits.  Recall that 
we normalize the simulated LMXB luminosity 
with the estimate of KW96 ($L_{X,tot} = 1.7 \times 10^{39}$ ergs s$^{-1}$), 
the HMXB population by the lower limit of Meurs \& van den Heuvel (1989) 
($N=28$ for $L_{X} > 10^{36} \, {\rm erg s^{-1}}$), our DNS 
formation rate by $10^{-6} {\rm yr}^{-1}$ (van den Heuvel 1995), and our 
derived retention fraction by 1\% for the globular cluster 
NGC 6397 (Drukier 1995).  We rule out only those pulsar velocity distributions 
whose Kolmogorov-Smirnov probability that the simulated and observed populations 
are derived from different parent populations is greater than 99\%.    

In \S 4.1 and 4.2, we present these ratios for a series of neutron-star 
kick distributions.  Since we are calculating upper limits, 
we require that all ratios be greater than unity for a successful 
kick distribution.  We use the results from \S 3 to create the best 
fit within the range of the many free binary-evolution parameters.  Unless 
otherwise noted, we use the standard set of assumptions described in \S 3:  $\alpha_
{IMF}=2.7$, mass limits $=10, 40 M_{\odot}$, $\alpha_{MR}=2.7$, $q_{0}=0.35$, 
$P(A_{0}) \propto 1/A_{0}$, $P(e_{0})=1.0$, $f_{trans}=0.5$, $\gamma = 2.1$, 
$\alpha_{CE}=1$, and $f_{wind}=1.0$.  

In \S 4.2, we use the results from our series of $\delta$-function distributions to 
derive the neutron-star kick distribution which best fits all of the observations.  
We find that double-peaked kick distributions best fit both the 
pulsar velocity data and the binary system formation rates and we study these 
distributions in more detail.  Our results depend most significantly on the 
distance measurements and we include a brief discussion of the effect of distance errors 
on our conclusions.  

\subsection{Maxwellian and Flat Distributions}

We ran a series of simulations with Maxwellian kick distributions 
for a variety of $v_{rms}$'s.  Figure 8 summarizes 
the results of these simulations, using the standard input 
parameters.  Table 3 gives the total number of bound neutron stars given 
the globular cluster models of Drukier (1995), along with  
the specific results for NGC 6397 and $\omega$ Cen.  The large kick velocities 
are required to explain the pulsar velocity distribution.  Over 20\% of the 
observed pulsars have transverse velocities greater than 500 km/s.  
Even without the effects of the galactic potential, high kick velocities 
are required to match the observations (Figure 4).   
We ran an alternate simulation using $q_{0}=0.15$ and the mass limits $=10,100 
M_{\odot}$.  The lower value for $q_{0}$ increases the number of LMXBs, while 
the higher mass limit 
allows more very massive stars to contribute to the HMXB and DNS populations.  The results 
are summarized in Figure 9 and Table 4 and can be directly compared to Figure 8 
and Table 3.  For this simulation, we set the binary parameters 
to maximize the production rate of the neutron-star populations.  Nevertheless, 
we see in Figure 9 that there is no acceptable solution.  
Keeping in mind that all of our ratios are upper limits, we conclude that 
it is impossible to fit the data with a Maxwellian kick distribution.  

Similarly, we ran a series of simulations with flat kick distributions ranging 
from a magnitude of 0 km/s to a maximum of $V_{max}$.  For these simulations, 
we used $q_{0}=0.15$ and the standard parameter set.  We see in Figure 10 and 
Table 5 that the fit is worse for this kick distribution than for a 
Maxwellian.  Again, we conclude that it is impossible to fit the data 
with a flat kick distribution.   

\subsection{Delta Function and Bimodal Distributions}

We repeated this set of simulations once again for single delta function kick 
distributions.  Figure 11 shows that we can find a delta function kick 
distribution that is consistent with the observed pulsar velocity 
distribution.  However, the delta function kick distribution fails to explain 
our entire data set, especially the globular cluster retention fraction (Table 6).  
The best fit to the pulsar 
data gives a kick distribution with a velocity near 500-600 km/s, 
higher than the mean pulsar velocity.  The mean kick must be higher due to the 
effects of binary evolution and the galactic potential which lower the mean 
neutron-star velocity after its initial kick.  As mentioned in \S 2, 
the galactic potential can alter the pulsar velocities by up to 40\% 
for some kick distributions.

Next, we use the results of the delta function simulations described 
above to infer a kick distribution that fits all observational 
constraints (LMXB, HMXB, DNS, globular cluster retention and the 
pulsar velocity distribution, which we bin into 5 roughly equal groups). 
To this effect, we approximate the kick velocity distribution by a weighted 
sum of individual delta functions of different kick magnitudes. 
We iterate on the weights until agreement between model and observations
is reached.

In practice, the kick velocity distribution is approximated by the 
sum of 7 individual delta functions with equally spaced values ranging from 0 to 
600 km/s. The observed pulsar velocity distribution fits into 5 
bins having approximately equal number of objects (we slightly underestimate 
the true velocities here by lumping the very high velocities into 
one bin of pulsars with velocities greater than 500 km/s). Therefore, our system
has 7 unknowns (the weights of the delta functions) and
9 constraints (3 binary systems, the globular cluster retention fraction,
and 5 pulsar velocity bins). For each realization, $\chi^2$ residuals 
are computed for each of the pulsar velocity bins and the best distribution
is the one that minimizes these residuals while fitting all the 
constraints.  The inclusion of additional delta function values did not lower 
the $\chi^2$ residuals, so our 7 delta functions represent the kick distribution 
sufficiently given the current data.  We varied the number of the bins of the observed 
pulsar velocity distribution as well as the binning procedure at fixed bin number and 
did not find any noticeable qualitative or quantitative change.  

The best fit to the entire data set obtained from this procedure is 
illustrated in Fig. 12. Notice that the best fit distribution has a
double-peaked profile\footnote{The best fit is actually trimodal.  
However, we do not have sufficient data to mandate this trimodal distribution 
and a bimodal kick distribution fits within the constraints of the data.  
Occam's razor limits us to a bimodal distribution at this time.}. 
This shape is required to explain all the 
observational constraints simultaneously. In other words, there must 
be a significant number of  neutron stars born with very small 
kicks in order to explain the low velocity population (binaries and 
cluster members), while pulsar velocities indicate that another significant 
fraction must receive appreciable kicks. This trend can
already be perceived in the raw data set; binary and galactic potential 
effects are not at the roots of this dichotomy.  We have fit the same 
distribution without the binary and cluster member constraints (thick 
line in Figure 12) and find that the low-velocity ($V_{kick} < 50$ km/s) 
population disappears.

To determine the robustness of this double-peaked distribution, we 
perform a number of tests.  First, we impose a lower limit to the 
weights of the intermediate bins (100 km/s - 400 km/s inclusive) 
and recompute the $\chi^2$ residuals. 
We see from figure 12 that the distribution not only tries to keep its 
double-peaked profile, but the $\chi^2$ residuals increase dramatically.

Second, to ascertain the effect of uncertainties in the radio pulsar 
distances on the double-peaked nature of the distribution, we artificially
scale down the observed velocities, since an overestimate of the 
distance translates directly into an overestimation of the velocity.
The $\chi^2$ residuals for these best-fitting distributions
are plotted in figure 13.  If we reduce the distance estimate by $25$\%, 
the $\chi^2$ residual for a flat kick distribution is only a factor of 5 
times larger than our best-fitting double-peaked kick distribution.  
Thus, if there exists systematic errors in the distance measurements which 
overestimate the pulsar distances by over $25$\%, our conclusions requiring 
a double-peaked kick distribution will not hold.  

We compare this double-peaked profile fit to our fits with the single-peak  
kick distributions by repeating the process 
used on the previous kick distributions.  We again normalize the binary systems 
to calculate upper limits on their birthrates.  For these simulations, we use 
$q_{0}=0.15$ and the standard parameter set.  The first series of simulations uses 
two $\delta$-function kick amplitudes.  Roughly 30
0 km/s and the remaining 70\% are given a non-zero kick.  We range this 
velocity from $500$ to $950$ km/s.  Note in figure 14 that over this series of 
simulations, the kick distributions satisfy our minimum requirements for an allowed kick 
distribution.  We also perform a second series of simulations using a kick distribution where 30\% of
the pulsars are given a kick of 0 km/s and the remaining 70\% have 
a flat distribution with a mean of $625$ km/s and a range in thickness (Figure 15).
In both cases, the range of bound neutron stars in globular clusters is stable (28\%-30\%), 
corresponding to $2800$ and $1000$ neutron stars retained in NGC 6397 and $\omega$ Cen, 
respectively.  

\section{Conclusions}

We have created a Monte Carlo code which simulates the binary evolution of 
massive stellar systems and includes HMXB, LMXB, DNS, and radio pulsar phases.  
This code also follows the motions of these systems in the galactic potential.  
In addition, we calculate the retention fraction of neutron star systems in 
globular clusters.  For this paper, we restricted our attention to the 
consequences of intrinsic kicks given to neutron stars at birth.  
First and foremost, a neutron star kick with a mean magnitude above 
$400$ km/s is required to explain the pulsar velocity data for all 
the kick distributions we study.  Figure 16 compares the observed transverse 
velocity data with the simulated transverse velocities for 
a Maxwellian kick distribution, a delta-function kick distribution and our 
double-peaked distribution (all with means above $400$ km/s).  The 
transverse velocities of a simulation without any neutron star kicks reveals 
the necessity of the kicks.  Figure 16 also illustrates our claim that with the 
radio pulsar observations alone, low--number statistics prevents us from constraining 
the kick distribution, beyond simply requiring a kick.  However, if we include the constraints 
placed upon the kick distribution from the binary populations and the globular 
cluster retention fraction, we can rule out many of the kick distributions 
appearing in the literature, including Maxwellian, flat, and $\delta$-function 
distributions (KW96, Brandt \& Podsiadlowski 1995).  Distributions which fit 
all of these constraints do exist, 
all of which are double-peaked.   To explain the birthrates of the 
neutron star binary populations, we 
derive that roughly 30\% of the neutron stars receive almost no kick.  
To explain the radio-pulsar velocity distribution, the remaining $\sim$70\% 
receive a large kick ($600-700$ km/s).  

Of course, there are many caveats to these conclusions.  Our results 
depend sensitively upon the pulsar velocity distribution.  If the pulsar 
distances and, hence, the velocities, are systematically lower by 25\%, 
a bimodal distribution is no longer necessary to explain the observations.  
However, it would require an extensive revision in the velocities to render some 
sort of neutron star kick unnecessary.  Also, we rely heavily upon the reasonableness 
of binary population-synthesis models.  Although we have studied the effects of many 
parameters and have calculated upper limits for all of our production rates, 
we can not eliminate the possibility that alternate models can explain 
the data.  For instance, Iben \& Tutukov (1996) explain the pulsar 
velocity distribution using the old distance model with no kick whatsoever 
by allowing only the highest-velocity neutron stars formed in binary evolution to 
become radio pulsars.  Scrutiny of Figure 4 reveals that this is not possible 
unless we remove mass loss from winds or unless we ignore the results from 
hydrodynamical models and allow systems to survive common-envelope 
phases with He-stars.  Even so, we require that only the 
fastest 1\% of the neutron stars are observed as pulsars to explain the 
pulsar velocities from the old distance model.  For the new distance 
model, this percentage becomes prohibitively small ($\lesssim 0.1$\%).  
We incorporate the effects of a wide range in binary parameters, so that 
unless the understanding of binary population synthesis is drastically 
altered (winds, common envelope evolution, etc.), our basic conclusions 
still hold.   In regimes where comparisons are possible, our results 
agree well with the models of KW96 and DS95.  

This bimodal kick distribution has direct implications for a variety of 
objects whose evolution may involve a neutron star.  To meet the isotropy 
requirements for gamma-ray bursts using accreting neutron stars in a galactic 
model.  Podsiadlowski, Rees, \& Ruderman (1995) 
require neutron star kicks 
upwards of $600-700$ km/s.  Our bimodal kick distribution results in 
$\sim$70\% of the neutron star population with these velocities (Compare 
the post-binary evolution velocities in Figure 5).  The 
bimodal distribution provides a natural break between the low-velocity 
neutron stars, which form X-ray binaries, and the high-velocity 
neutron stars which might make up the gamma-ray burst population in a 
galactic model
(Leonard \& Colgate 1994).  

Our kick distribution can also be applied to explain O/B runaway stars.  
Although O/B runaway stars are not observed to be in close binaries, 
the observations do not preclude wide binary systems.  (Gies \& Bolton 
1986, Sayer, Nice \& Kaspi 1996, Philp {\it et al.} 1996).  Figure 17 plots 
the distribution of velocities of O/B stars, both bound and unbound, 
assuming no neutron star kick.  The unbound O/B stars are all moving slower 
than $50$ km/s.  The bound systems have significantly higher 
velocities, but very few O/B stars have velocities greater than $100$ km/s.  
However, using our bimodal kick distribution, we see that 
unbound O/B stars can achieve velocities in excess of $200$ km/s.  

This bimodal distribution poses an additional problem for kick mechanisms.  
Not only must a kick mechanism produce neutron stars with velocities 
greater than 500 km/s, but the mechanism must be ineffective for 
a subset of the neutron star population.  Since the submission of this 
paper, work by Cordes \& Chernoff (1997) has appeared in the literature.  
This work concentrates upon an understanding of the biases of the 
pulsar velocity data and estimates of the pulsar ages.  It predicts 
a double-peaked distribution of the {\it pulsar velocity distribution}, 
qualitatively agreeing with our results of the kick distribution.  
However, the quantitative differences make it clear that much more 
work must be done to gain a definitive answer on kick velocities.  

Our simulations can be seen as the first step in constraining the natal 
neutron-star kick distribution.  Although the double-peaked nature of the 
kick distribution is required by our calculations, there remains a wealth 
of observational data which can be used to constrain the quantitative nature 
and individual structure of the peaks.   The orbital characteristics of the 
binary systems formed using a variety of kick distributions may also provide 
insight into the specifics of the  distribution.  For instance, wide-orbit LMXBs 
(KW96) and short-period DNSs (Fryer \& Kalogera 1997) may further constrain the kick 
distribution.  We have not yet explored variations in the galactic potential and the scale-height 
distributions of the various neutron star systems.  With improved distances 
and with an increasing sample of radio pulsars, we hope to apply this technique not 
just to constrain the neutron star kick and binary evolution, but also the galactic 
potential.  

\acknowledgements
It is a pleasure to thank Vicky Kalogera for many helpful discussions on 
binary population synthesis and comments on the manuscript.  We are grateful 
to Chad Engelbracht for many helpful discussions which clarified much of 
the paper.  Paul Harding also provided useful advice on galaxy dynamics and 
stellar populations.  We would also like to thank an the referee, Rachel 
Dewey,  for many helpful comments and suggestions which improved 
the paper immensely.  The work of C.F. and W.B. was partially supported 
by NSF grant AST 9206738 and a ``Profil 2'' grant from the Swiss 
National Science Foundation.  A.B. acknowledges support from NSF grant 
AST 9217322.

\begin{deluxetable}{lccc}
\tablewidth{35pc}
\tablecaption{Parameters:  Initial Conditions\tablenotemark{a}}
\tablehead{ & \colhead{LMXB} & \colhead{HMXB} & \colhead{DNS} \\
 & \colhead{$10^{39}$ erg s$^{-1}$} & 
\colhead{N with $L_{X}>10^{36}$ erg s$^{-1}$}      
& \colhead{Rate ($10^{-6} {\rm y}^{-1}$)}}

\startdata
$V_{kick}=0 \, {\rm km \, s}^{-1}$ & & & \nl
``standard'' & 2.80 & 92.7 & 45.2 \nl
$\alpha_{IMF}=2.1$ & 2.62 & 142.4 & 62.7 \nl
$M_{min,max}=8,40$ & 1.36 & 65.8 & 26.0 \nl
$M_{min,max}=10,100$ & 1.68 & 47.7 & 36.1 \nl
$\alpha_{MR}=1.0$\tablenotemark{b} & 0.227 & 60.9 & 200. \nl 
$\alpha_{MR}=0.0$\tablenotemark{b} & 2.41 & 82.6 & 135. \nl 
$\alpha_{MR}=2.7, q_0=0.15$\tablenotemark{c} & 11.1 & 152.1 & 26.8 \nl
$P(e)=\delta(e)$ & 2.91 & 71.2 & 54.6 \nl
$A_{max}=10^6$ & 1.26 & 48.2 & 73.4 \nl

& & & \nl
$V_{kick}=200 \, {\rm km \, s}^{-1}$ & & & \nl
``standard'' & 4.16 & 40.3 & 1.15 \nl
$\alpha_{IMF}=2.1$ & 4.47 & 53.3 & 1.42 \nl
$M_{min,max}=8,40$ & 2.81 & 35.5 & 0.641 \nl
$M_{min,max}=10,100$ & 2.56 & 29.4 & 0.820 \nl
$\alpha_{MR}=1.0$ & 0.420 & 35.1 & 3.56 \nl 
$\alpha_{MR}=0.0$ & 3.07 & 42.3 & 2.49 \nl 
$\alpha_{MR}=2.7, q_0=.15$ & 19.3 & 100.5 & 0.751 \nl
$P(e)=\delta(e)$ & 5.12 & 32.8 & 1.11 \nl
$A_{max}=10^6$ & 1.71 & 21.8 & 0.547 \nl

& & & \nl
$V_{kick}=400 \, {\rm km \, s}^{-1}$ & & & \nl
``standard'' & 1.68 & 13.0 & 0.474 \nl
$\alpha_{IMF}=2.1$ & 1.98 & 16.9 & 0.656 \nl
$M_{min,max}=8,40$ & 1.24 & 10.7 & 0.289 \nl
$M_{min,max}=10,100$ & 0.937 & 10.5 & 0.384 \nl
$\alpha_{MR}=1.0$ & 0.159 & 10.6 & 1.85 \nl 
$\alpha_{MR}=2.7, q_0=0.35 $ & 1.68 & 13.0 & 0.474 \nl
$\alpha_{MR}=2.7, q_0=.15$ & 6.95 & 29.8 & 0.437 \nl
$P(e)=\delta(e)$ & 1.96 & 10.7 & 0.491 \nl
$A_{max}=10^6$ & 0.660 & 7.51 & 0.265 \nl

\tablenotetext{a}{For these simulations, we use a ``standard'' set of 
parameters:  $\alpha_{IMF}=2.7$, mass limits $=10,40 M_{\odot}$, 
$\alpha_{MR}=2.7$, $q_{0}=0.35$, $P(A_{0}) \propto 1/A_{0}$, 
$P(e_{0})=1.0$, $f_{trans}=0.5$, $\gamma = 2.1$, $\alpha_{CE}=1$, 
and $f_{wind}=1.0$.  We have 
combined the simulations with a SN rate\,$=0.01 {\rm y}^{-1}$ and a 
binary fraction determined by the choice of mass ratio (unless 
otherwise stated, we use $0.35$).  We use the technique in \S 
2 to determine each population.}

\tablenotetext{b}{For $\alpha_{MR}=1.0,0.0$, we assume a binary 
fraction of 0.43.}

\tablenotetext{c}{For $\alpha_{MR}=2.7,\,q_o=0.15$, we assume a binary 
fraction of 0.65.}

\enddata
\end{deluxetable}

\begin{deluxetable}{lccc}
\tablewidth{35pc}
\tablecaption{Parameters:  Binary Evolution\tablenotemark{a}}
\tablehead{ & \colhead{LMXB} & \colhead{HMXB} & \colhead{DNS} \\
 & \colhead{$10^{39}$ erg s$^{-1}$} & 
\colhead{N with $L_{X}>10^{36}$ erg s$^{-1}$}      
& \colhead{Rate ($10^{-6} {\rm y}^{-1}$)}}

\startdata
$V_{kick}=0 \, {\rm km \, s}^{-1}$ & & & \nl
``standard'' & 2.80 & 92.7 & 45.2 \nl
$f_{trans}=0.1$ & 2.47 & 60.3 & 45.0 \nl
$f_{trans}=0.9$ & 2.39 & 61.9 & 36.0 \nl
$\alpha_{CE}=0.2$ & $\lesssim 0.028$ & 4.82 & 45.2 \nl
$\alpha_{CE}=2.0$ & 4.94 & 181.5 & 45.3 \nl
$f_{wind}=0$ & 1.75 & 89.7 & 39.2 \nl
$\gamma=1.5$ & 2.48 & 238.2 & 36.3 \nl 
& & & \nl
$V_{kick}=200 \, {\rm km \, s}^{-1}$ & & & \nl
``standard'' & 4.16 & 40.3 & 1.15 \nl
$f_{trans}=0.1$ & 4.06 & 23.3 & 2.00 \nl
$f_{trans}=0.9$ & 4.43 & 32.0 & 0.955 \nl
$\alpha_{CE}=0.2$ & $\lesssim 0.028$ & 2.28 & 0.0232 \nl
$\alpha_{CE}=2.0$ & 11.4 & 71.7 & 2.13 \nl
$f_{wind}=0$ & 24.2 & 65.0 & 6.01 \nl
$\gamma=1.5$ & 4.71 & 85.9 & 1.38 \nl 
& & & \nl
$V_{kick}=400 \, {\rm km \, s}^{-1}$ & & & \nl
``standard'' & 1.68 & 13.0 & 0.474 \nl
$f_{trans}=0.1$ & 1.54 & 6.94 & 0.854 \nl
$f_{trans}=0.9$ & 1.71 & 9.93 & 0.513 \nl
$\alpha_{CE}=0.2$ & $\lesssim 0.028$ & 0.839 & 0.00664 \nl
$\alpha_{CE}=2.0$ & 3.7 & 20.4 & 0.927 \nl
$f_{wind}=0$ & 13.3 & 21.3 & 2.43 \nl
$\gamma=1.5$ & 1.45 & 29.3 & 0.577 \nl 

\tablenotetext{a}{For these simulations, we use a ``standard'' set of 
parameters:  $\alpha_{IMF}=2.7$, mass limits $=10,40 M_{\odot}$, 
$\alpha_{MR}=2.7$, $q_{0}=0.35$, $P(A_{0}) \propto 1/A_{0}$, 
$P(e_{0})=1.0$, $f_{trans}=0.5$, $\gamma = 2.1$, $\alpha_{CE}=1$, 
and $f_{wind}=1.0$.  We have combined the simulations with a SN 
rate\,$=0.01 {\rm y}^{-1}$ and a 
binary fraction determined by the choice of mass ratio (unless 
otherwise stated, we use $0.35$).  We use the technique in \S 
2 to determine each population.}

\enddata
\end{deluxetable}

\begin{deluxetable}{lccccc}
\tablewidth{40pc}
\tablecaption{NS Retention:  Maxwellian Kick Distribution:  $q_0=0.35$\tablenotemark{a}}
\tablehead{ $ (v_{rms}^{2})^{1/2}$ 
 & \colhead{Range in RF\tablenotemark{b}} & \colhead{RF} & 
\colhead{NS retained} & \colhead{RF} &
\colhead{NS retained} \\
$(\rm{km} \; {\rm s}^{-1}) $& & \multicolumn{2}{c}{NGC6397} & \multicolumn{2}{c}{$\omega$ Cen}}

\startdata
$0.0$ & $91.6-99.9$\% & $98$\% & $9.5\times 10^{3}$ & $99.1$\% & $3.5\times 10^{3}$ \nl
$50.0$ & $4.71-49.0$\% & $15.4$\% & $1.5\times 10^{3}$ & $29.2$\% & $1.0\times 10^{3}$ \nl
$100.0$ & $1.91-38.3$\% & $5.30$\% & $5.1\times 10^{2}$ & $11.1$\% & $3.9\times 10^{2}$ \nl
$150.0$ & $1.19-37.5$\% & $2.95$\% & $2.9\times 10^{2}$ & $5.83$\% & $2.0\times 10^{2}$ \nl
$200.0$ & $0.88-18.9$\% & $2.06$\% & $2.0\times 10^{2}$ & $3.88$\% & $1.4\times 10^{2}$ \nl
$250.0$ & $0.70-13.6$\% & $1.60$\% & $1.6\times 10^{2}$ & $2.89$\% & $1.0\times 10^{2}$ \nl
$300.0$ & $0.57-10.2$\% & $1.27$\% & $1.2\times10^{2}$ & $2.26$\% & $79$ \nl
$350.0$ & $0.49-8.1$\% & $1.08$\% & $1.0\times10^{2}$ & $1.90$\% & $67$ \nl
$400.0$ & $0.42-6.6$\% & $0.93$\% & $90$ & $1.63$\% & $57$ \nl
$450.0$ & $0.38-5.5$\% & $0.83$\% & $81$ & $1.42$\% & $49$ \nl
$500.0$ & $0.34-4.7$\% & $0.74$\% & $72$ & $1.28$\% & $46$ \nl
$550.0$ & $0.31-4.1$\% & $0.66$\% & $64$ & $1.12$\% & $37$ \nl
$600.0$ & $0.29-3.6$\% & $0.61$\% & $59$ & $1.02$\% & $35$ \nl

\tablenotetext{a}{Standard Parameters: $\alpha_{IMF}=2.7$, 
mass limits $=10,40 M_{\odot}$, $\alpha_{MR}=2.7$, $q_{0}=0.35$, 
$P(A_{0}) \propto 1/A_{0}$, $P(e_{0})=1.0$, $f_{trans}=0.5$, 
$\gamma = 2.1$, $\alpha_{CE}=1$, and $f_{wind}=1.0$.}
\tablenotetext{b}{The range in retention fractions (RF)  
is determined using the Fokker-Planck models of 
Drukier (1995).}

\enddata
\end{deluxetable}

\begin{deluxetable}{lccccc}
\tablewidth{40pc}
\tablecaption{NS Retention:  Maxwellian Kick Distribution: $q_0=0.15$\tablenotemark{a}}
\tablehead{ $ (v_{rms}^{2})^{1/2}$ 
 & \colhead{Range in RF} & \colhead{RF} & 
\colhead{NS retained} & \colhead{RF} &
\colhead{NS retained} \\
$(\rm{km} \; {\rm s}^{-1}) $& & \multicolumn{2}{c}{NGC6397} & 
\multicolumn{2}{c}{$\omega$ Cen}}

\startdata
$0.0$ & $89.0-99.7$\% & $95$\% & $9.2\times 10^{3}$ & $97$\% & $3.4\times 10^{3}$ \nl
$50.0$ & $3.96-70.4$\% & $18.4$\% & $1.8\times 10^{3}$ & $39.6$\% & $1.4\times 10^{3}$ \nl
$100.0$ & $1.25-54.7$\% & $4.8$\% & $4.7\times 10^{2}$ & $12.7$\% & $4.4\times 10^{2}$ \nl
$150.0$ & $0.70-37.5$\% & $2.25$\% & $2.2\times 10^{2}$ & $5.65$\% & $2.0\times 10^{2}$ \nl
$200.0$ & $0.50-25.0$\% & $1.41$\% & $1.4\times 10^{2}$ & $3.26$\% & $1.2\times 10^{2}$ \nl
$250.0$ & $0.41-17.0$\% & $1.04$\% & $1.0\times 10^{2}$ & $2.21$\% & $77$ \nl
$300.0$ & $0.31-12.1$\% & $0.77$\% & $75$ & $1.59$\% & $56$ \nl
$350.0$ & $0.28-8.9$\% & $0.67$\% & $65$ & $1.3$\% & $46$ \nl
$400.0$ & $0.23-6.9$\% & $0.55$\% & $53$ & $1.1$\% & $37$ \nl
$450.0$ & $0.21-5.4$\% & $0.47$\% & $46$ & $0.89$\% & $31$ \nl
$500.0$ & $0.19-4.4$\% & $0.44$\% & $43$ & $0.80$\% & $28$ \nl
$550.0$ & $0.18-3.7$\% & $0.39$\% & $38$ & $0.69$\% & $24$ \nl
$600.0$ & $0.17-2.4$\% & $0.34$\% & $33$ & $0.61$\% & $21$ \nl

\tablenotetext{a}{same as table 3, $q_{0}=0.15$.} 

\enddata
\end{deluxetable}

\begin{deluxetable}{lccccc}
\tablewidth{40pc}
\tablecaption{NS Retention:  Flat Kick Distribution\tablenotemark{a}}
\tablehead{ $ Mean Velocity $ 
 & \colhead{Range in RF\tablenotemark{b}} & \colhead{RF} & 
\colhead{NS retained} & \colhead{RF} &
\colhead{NS retained} \\
$(\rm{km} \; {\rm s}^{-1}) $& & \multicolumn{2}{c}{NGC6397} & \multicolumn{2}{c}{$\omega$ Cen}}

\startdata
$50.0$ & $9.44-69.9$\% & $20.1$\% & $1.9\times 10^{3}$ & $33.3$\% & $1.2\times 10^{3}$ \nl
$150.0$ & $3.13-32.1$\% & $6.66$\% & $646$ & $11.0$\% & $385$ \nl
$200.0$ & $2.35-24.1$\% & $5.00$\% & $485$ & $8.28$\% & $290$ \nl
$250.0$ & $1.88-19.2$\% & $3.99$\% & $387$ & $6.62$\% & $232$ \nl
$300.0$ & $1.56-16.0$\% & $3.33$\% & $323$ & $5.52$\% & $193$ \nl
$350.0$ & $1.34-13.7$\% & $2.85$\% & $276$ & $4.73$\% & $166$ \nl

\tablenotetext{a}{Parameters: $\alpha_{IMF}=2.7$, 
mass limits $=10,40 M_{\odot}$, $\alpha_{MR}=2.7$, $q_{0}=0.15$, 
$P(A_{0}) \propto 1/A_{0}$, $P(e_{0})=\delta(e_{0})$, $f_{trans}=0.5$, 
$\gamma = 2.1$, $\alpha_{CE}=1$, and $f_{wind}=1.0$.}
\tablenotetext{b}{The range in retention fractions (RF) 
is determined using the Fokker-Planck models of  
Drukier (1995).}

\enddata
\end{deluxetable}

\begin{deluxetable}{lccccc}
\tablewidth{40pc}
\tablecaption{NS Retention:  $\delta$-Function Kick Distribution\tablenotemark{a}}
\tablehead{ $ Velocity $ 
 & \colhead{Range in RF\tablenotemark{b}} & \colhead{RF} & 
\colhead{NS retained} & \colhead{RF} &
\colhead{NS retained} \\
$(\rm{km} \; {\rm s}^{-1}) $& & \multicolumn{2}{c}{NGC6397} & \multicolumn{2}{c}{$\omega$ Cen}}

\startdata
$0.0$ & $89.5-99.9$\% & $97.1$\% & $9.4\times 10^{3}$ & $98.7$\% & $3.5\times 10^{3}$ \nl
$50.0$ & $0.19-96.8$\% & $3.62$\% & $3.5\times 10^{2}$ & $42.5$\% & $1.5\times 10^{3}$ \nl
$100.0$ & $.011-67.3$\% & $0.077$\% & $7$ & $0.398$\% & $13.9$ \nl
$150.0$ & $4.2\times10^{-3}-36.6$\% & $0.033$\% & $3$ & $0.127$\% & $4$ \nl
$200.0$ & $2.8\times10^{-3}-11.8$\% & $0.028$\% & $3$ & $0.078$\% & $3$ \nl
$250.0$ & $1.2\times10^{-3}-1.39$\% & $0.016$\% & $2$ & $0.045$\% & $2$ \nl
$300.0$ & $5.1\times10^{-4}-0.43$\% & $7.6\times10^{-3}$\% & $1$ & $0.025$\% & $1$ \nl
$350.0$ & $2.2\times10^{-5}-0.30$\% & $5.5\times10^{-3}$\% & $1$ & $0.021$\% & $1$ \nl
$400.0$ & $1.7\times10^{-6}-0.21$\% & $2.1\times10^{-3}$\% & $0$ & $0.011$\% & $0$ \nl
$450.0$ & $0.0-0.16$\% & $1.0\times10^{-3}$\% & $0$ & $7.9\times10^{-3}$\% & $0$ \nl
$500.0$ & $0.0-0.11$\% & $7.1\times10^{-4}$\% & $0$ & $4.7\times10^{-3}$\% & $0$ \nl
$550.0$ & $0.0-0.080$\% & $1.7\times10^{-4}$\% & $0$ & $2.3\times10^{-3}$\% & $0$ \nl
$600.0$ & $0.0-0.062$\% & $7.9\times10^{-5}$\% & $0$ & $1.4\times10^{-3}$\% & $0$ \nl
$650.0$ & $0.0-0.023$\% & $5.1\times10^{-6}$\% & $0$ & $2.5\times10^{-4}$\% & $0$ \nl

\tablenotetext{a}{Standard Parameters: $\alpha_{IMF}=2.7$, 
mass limits $=10,40 M_{\odot}$, $\alpha_{MR}=2.7$, $q_{0}=0.35$, 
$P(A_{0}) \propto 1/A_{0}$, $P(e_{0})=1.0$, $f_{trans}=0.5$, 
$\gamma = 2.1$, $\alpha_{CE}=1$, and $f_{wind}=1.0$.}
\tablenotetext{b}{The range in retention fractions (RF) 
is determined using the Fokker-Planck models of  
Drukier (1995).}

\enddata
\end{deluxetable}


\clearpage

\begin{figure}
\plotfiddle{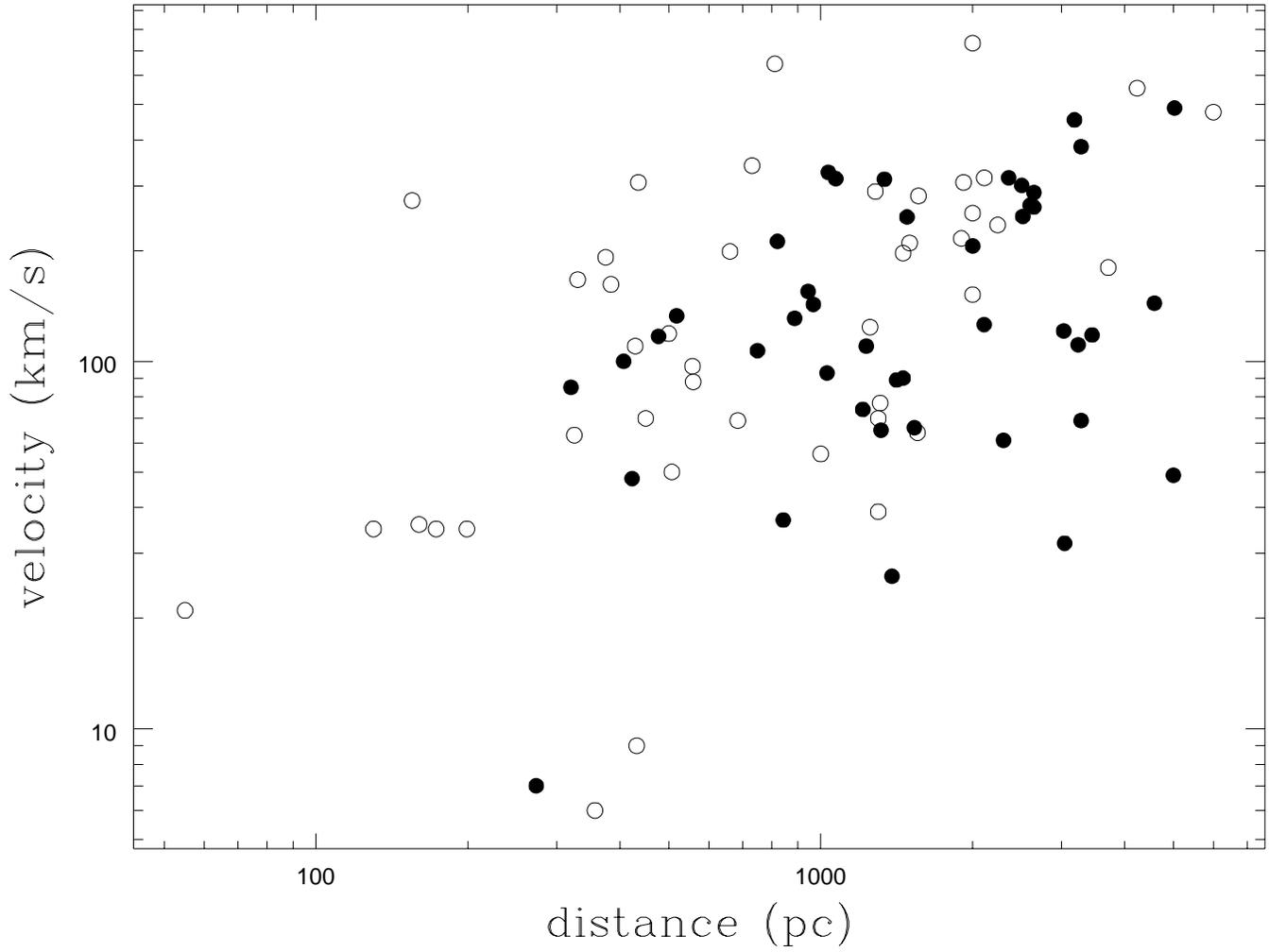}{7in}{-90}{70}{70}{-280}{520}
\caption{The observed transverse velocities of radio pulsars versus 
their distances D from the Sun (Harrison {\it et al.} 1993).  The open 
circles denote the 44 pulsars with proper motions determined by 
Harrison {\it et al.} and the filled circles are the 43 additional pulsars 
with proper motions calculated by other techniques.}
\end{figure}

\begin{figure}
\plotfiddle{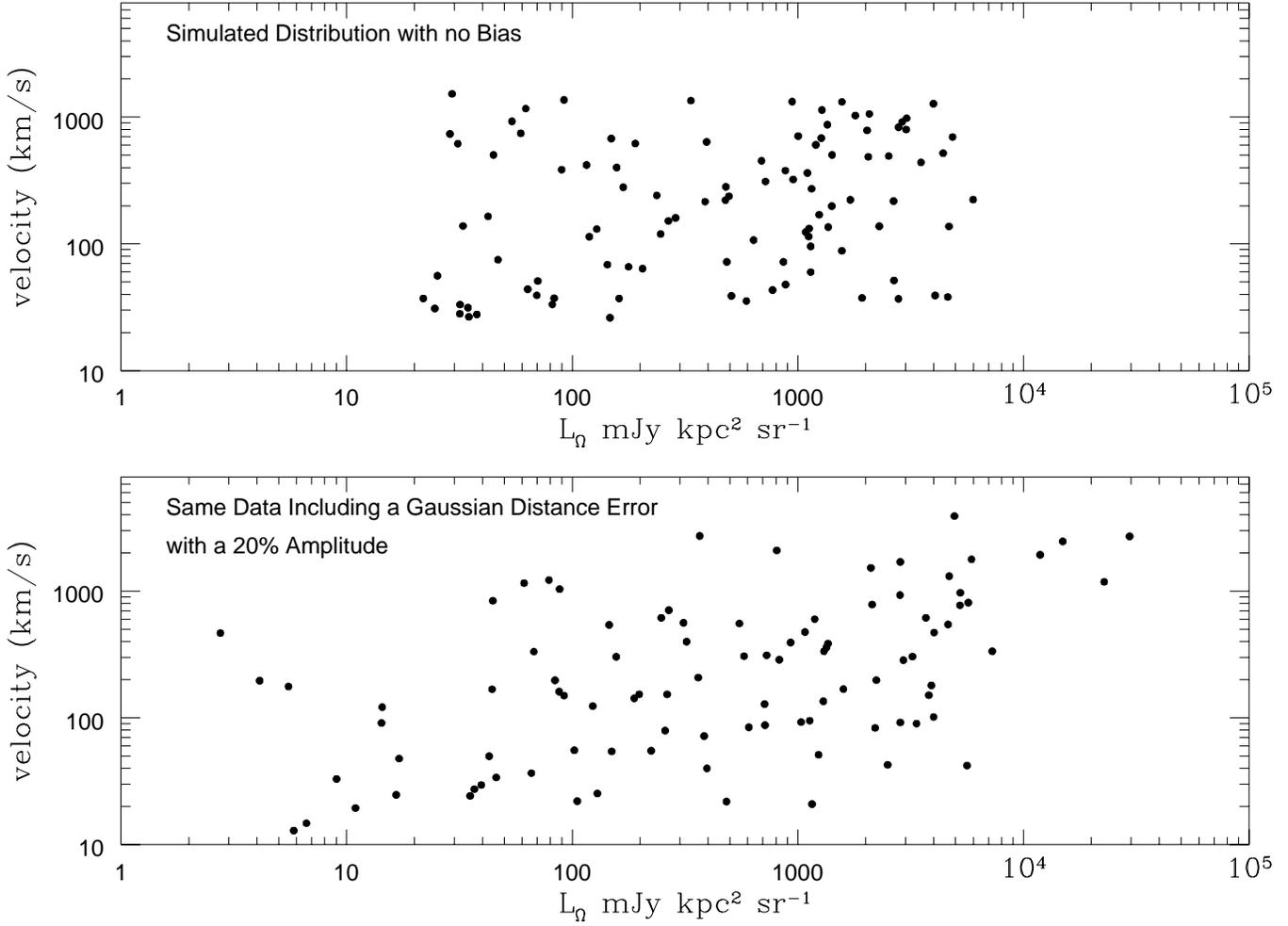}{7in}{-90}{70}{70}{-280}{520}
\caption{Luminosity versus velocity for a simulated sample (no biases) without 
distance errors on the left and with distance errors on the right.  The 
pulsars in the simulated sample are evenly chosen in velocity/luminosity 
space with no intrinsic biases.  We assume 
the distance errors are Gaussian with a magnitude of 20\% distance 
(Taylor \& Cordes 1993 assume gaussian distance errors on the order of 10\%).  The 
higher-velocity pulsars {\it appear} to be more luminous.}
\end{figure}

\begin{figure}
\plotfiddle{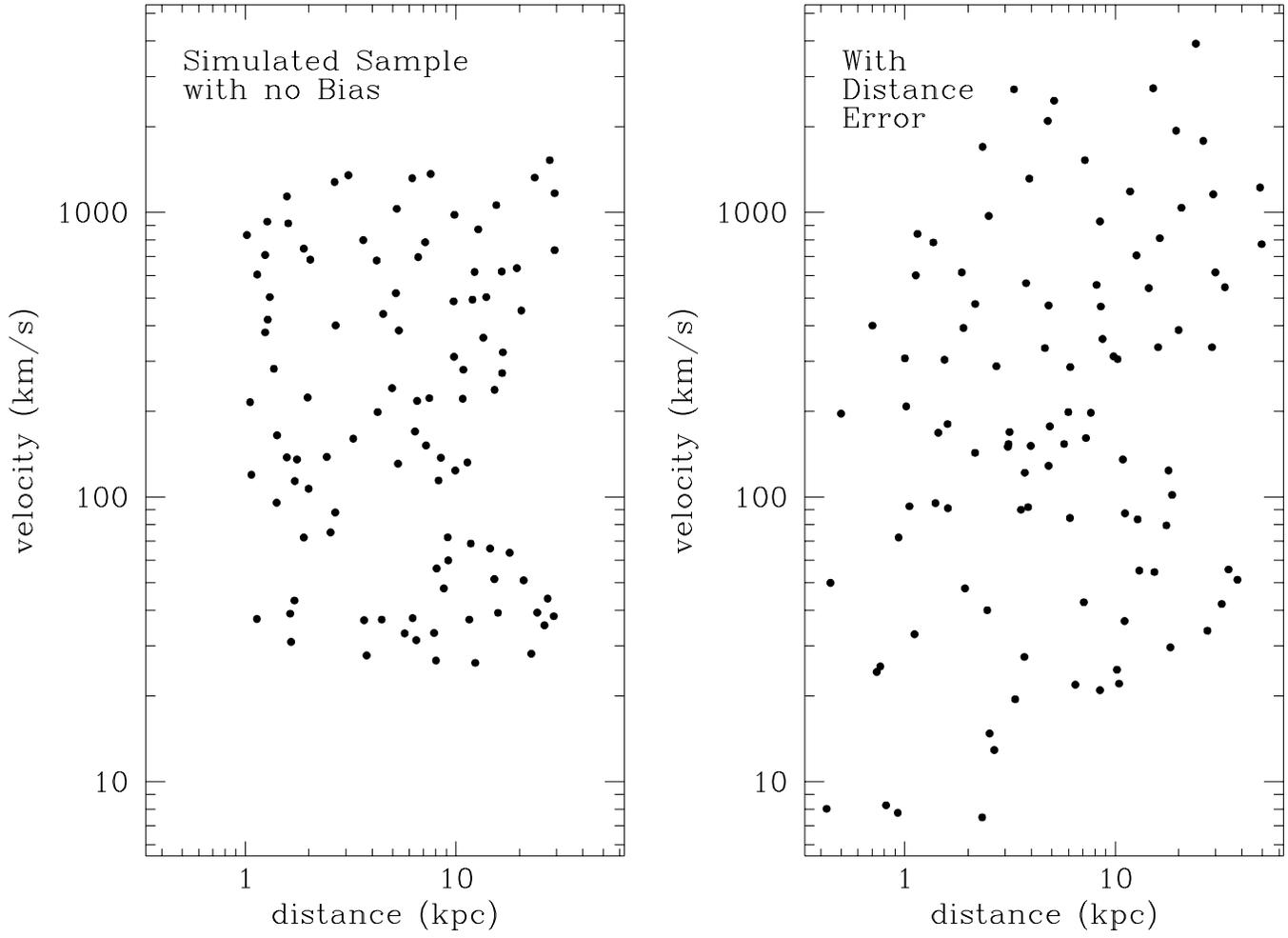}{7in}{-90}{70}{70}{-280}{520}
\caption{The same as Figure 2, except velocity is plotted versus distance.  
Note that with distance errors, another fictitious bias appears with the 
nearby pulsars tending to have lower velocities.}
\end{figure}

\begin{figure}
\plotfiddle{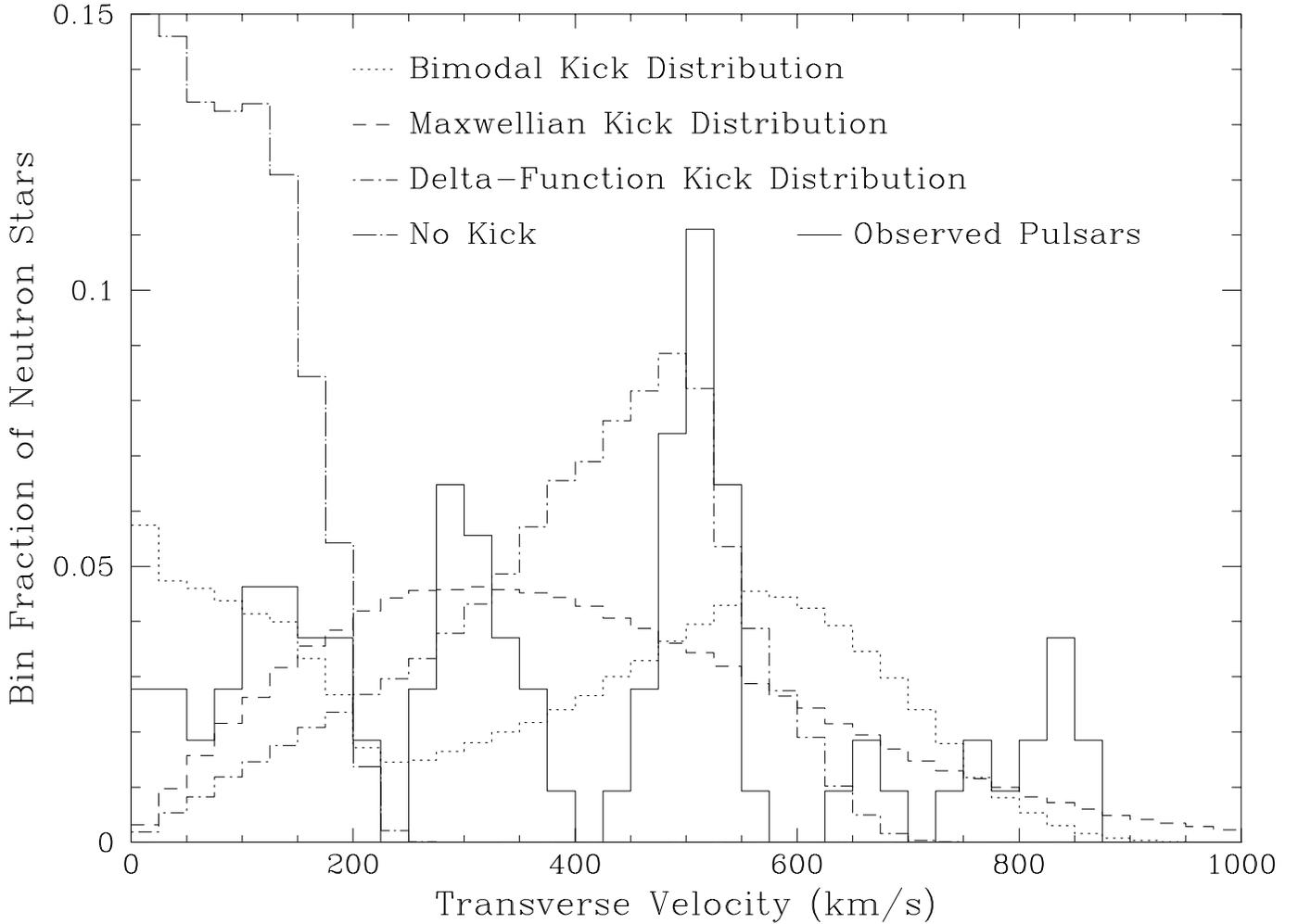}{6in}{-90}{70}{70}{-280}{450}
\caption{Transverse velocity distributions for the best fitting 
double-peaked, Maxwellian, and delta-function kick distributions along 
with a smoothed (within the quoted errors - Taylor, Manchester, \& Lyne 1993) 
distribution of the observed pulsar transverse velocities.  
Included for comparison is the transverse velocity distribution for 
a simulation with no natal neutron star kick.  The simulation with 
no neutron star kicks clearly does not fit the data, but the other 
distributions can not be ruled out with such small number statistics.}
\end{figure}

\begin{figure}
\plotfiddle{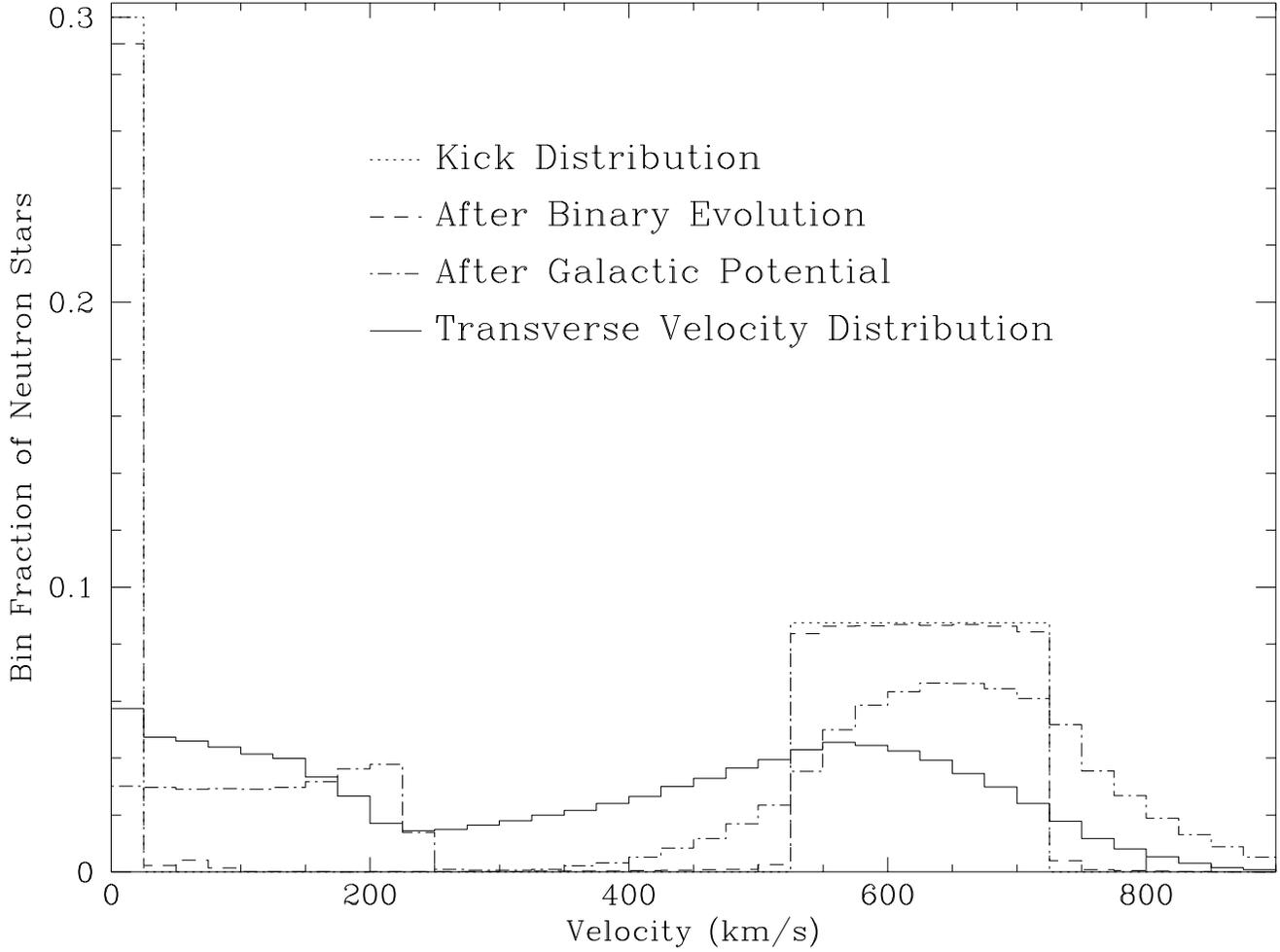}{7in}{-90}{70}{70}{-280}{520}
\caption{The dependence of the pulsar velocity distribution for a 
double-peaked kick distribution (dotted curve) on binary evolution 
(dashed curve) and the galactic potential (dot-dashed curve).  The 
solid curve gives the pulsar transverse-velocity distribution which 
is then compared to the observations.  Note that the transverse velocity 
distribution bears little resemblance to its parent kick distribution, 
illustrating the importance of including the effects of binary evolution 
and motion in the galaxy in extracting the actual kick distribution from 
the observations.}
\end{figure} 

\begin{figure}
\plotfiddle{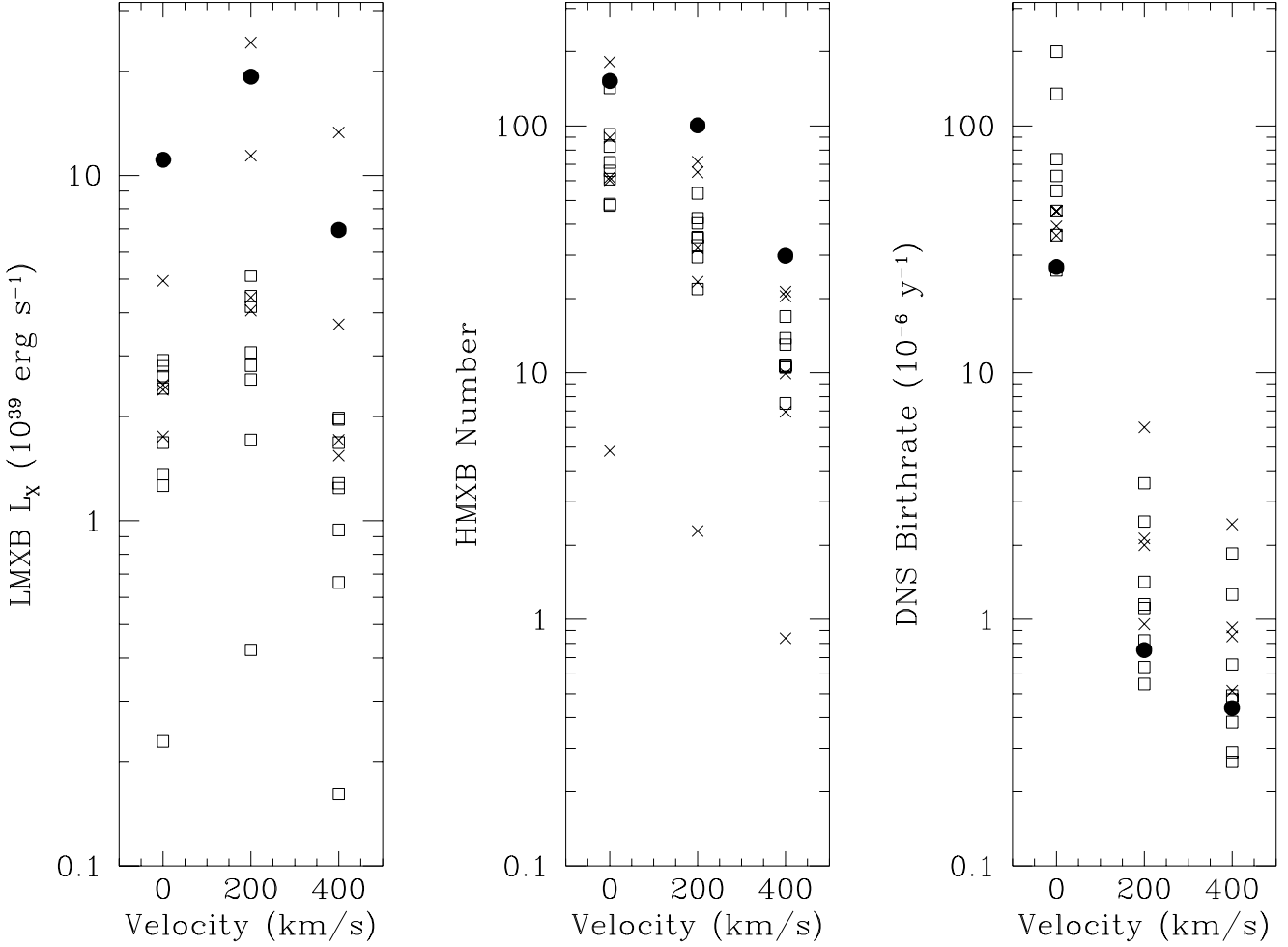}{7in}{-90}{70}{70}{-280}{520}
\caption{The dependence of the various neutron star binaries with respect 
to the various initial binary and evolutionary parameters and velocity. The 
filled circles correspond to the standard parameter set.  The open squares 
correspond to the initial binary parameters (\S 3.1) and the crosses 
correspond to the binary evolution parameters (\S 3.2).}
\end{figure}

\begin{figure}
\plotfiddle{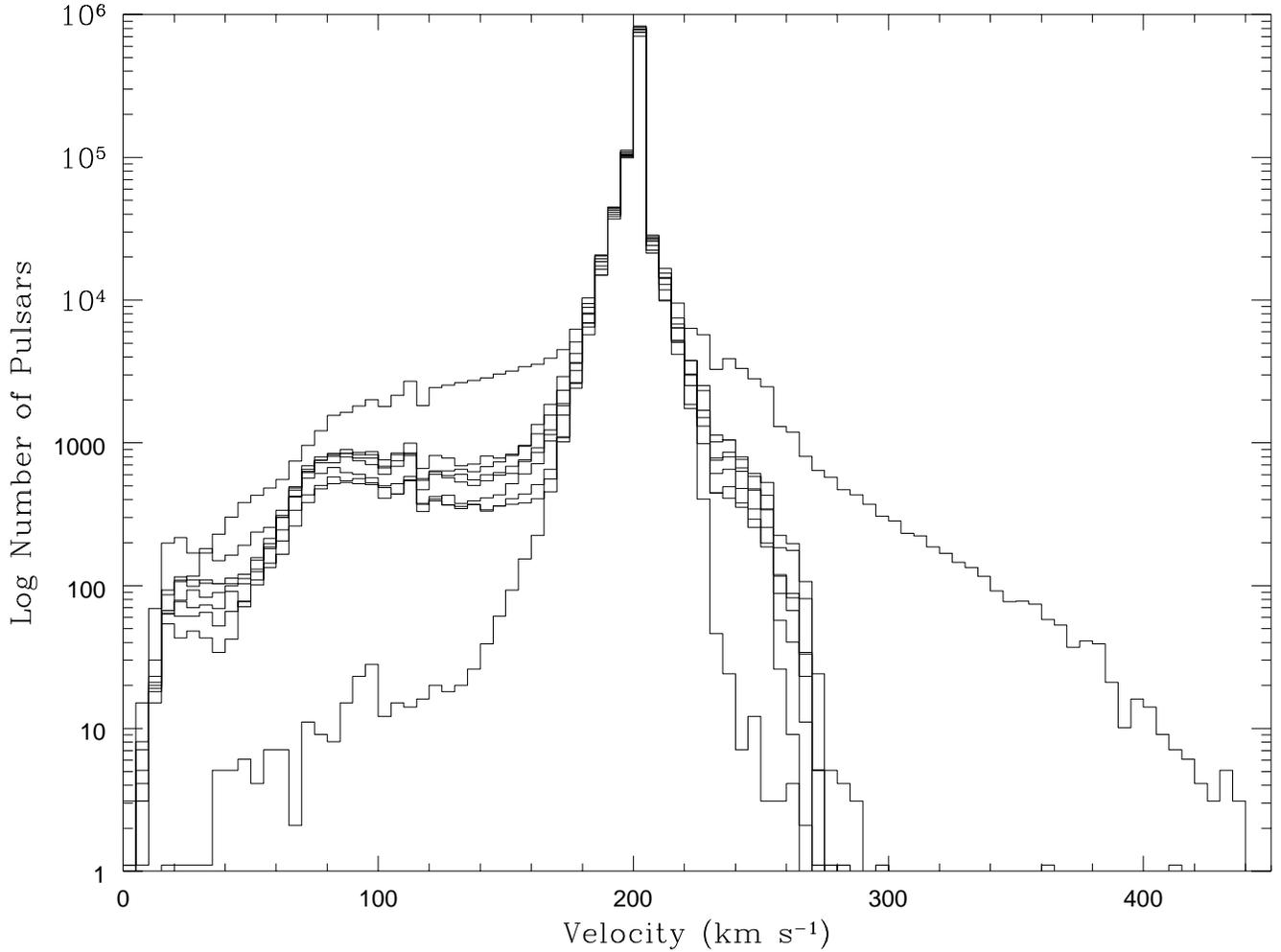}{7in}{-90}{70}{70}{-280}{520}
\caption{The three-dimensional velocity distribution of pulsars for a range of initial and 
binary-evolution parameters with a $\delta$-function kick of 200 km/s.  
The binary parameters affect only a small number of the 
pulsars.  The typical pulsar velocity is unchanged by even extreme changes 
in the binary parameters.  The largest differences come from assuming that 
there is no mass lost from winds which allows very tight pre-SN systems to form 
and leads to the broadest profile.  The narrow profile is the velocity distribution 
calculated by setting $\alpha_{CE}=0.2$ which prevents the formation of these 
close binaries.  Note, however, that the bulk of the binary parameters have 
little effect on the pulsar velocity distribution.}
\end{figure}

\begin{figure}
\plotfiddle{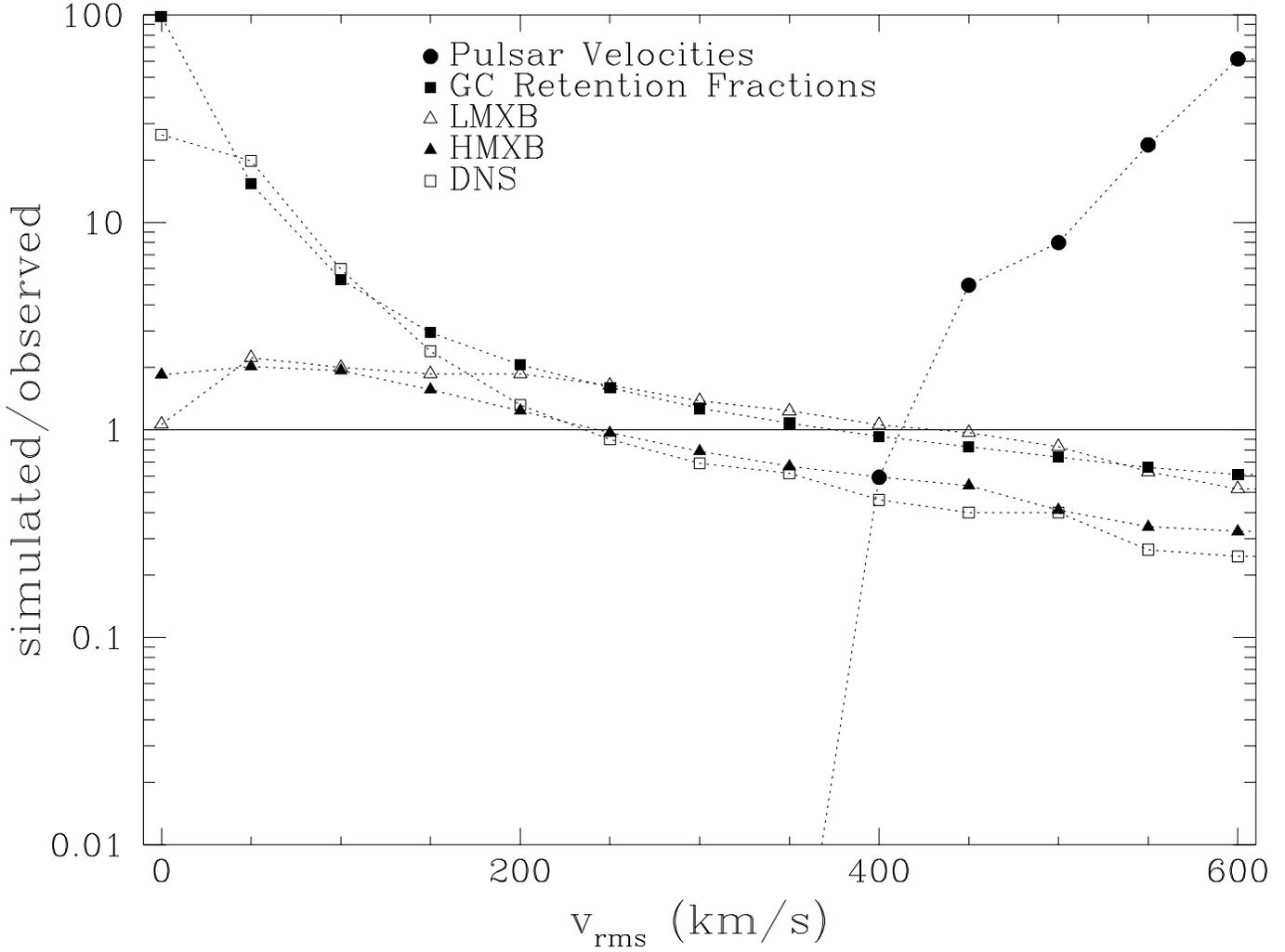}{7in}{-90}{70}{70}{-280}{520}
\caption{The simulated populations normalized by the observations versus the 
root-mean-square velocity for a Maxwellian Distribution.  For the pulsars, 
we plot $(99.99-P)$ where $P$ is the percentage probability that the 
simulated velocity distribution and the observed velocity distribution 
are not from the same parent population.  We use our standard set of binary 
parameters.  A ``successful'' solution is one for which all the normalized numbers 
are greater than unity simultaneously.}
\end{figure}

\begin{figure}
\plotfiddle{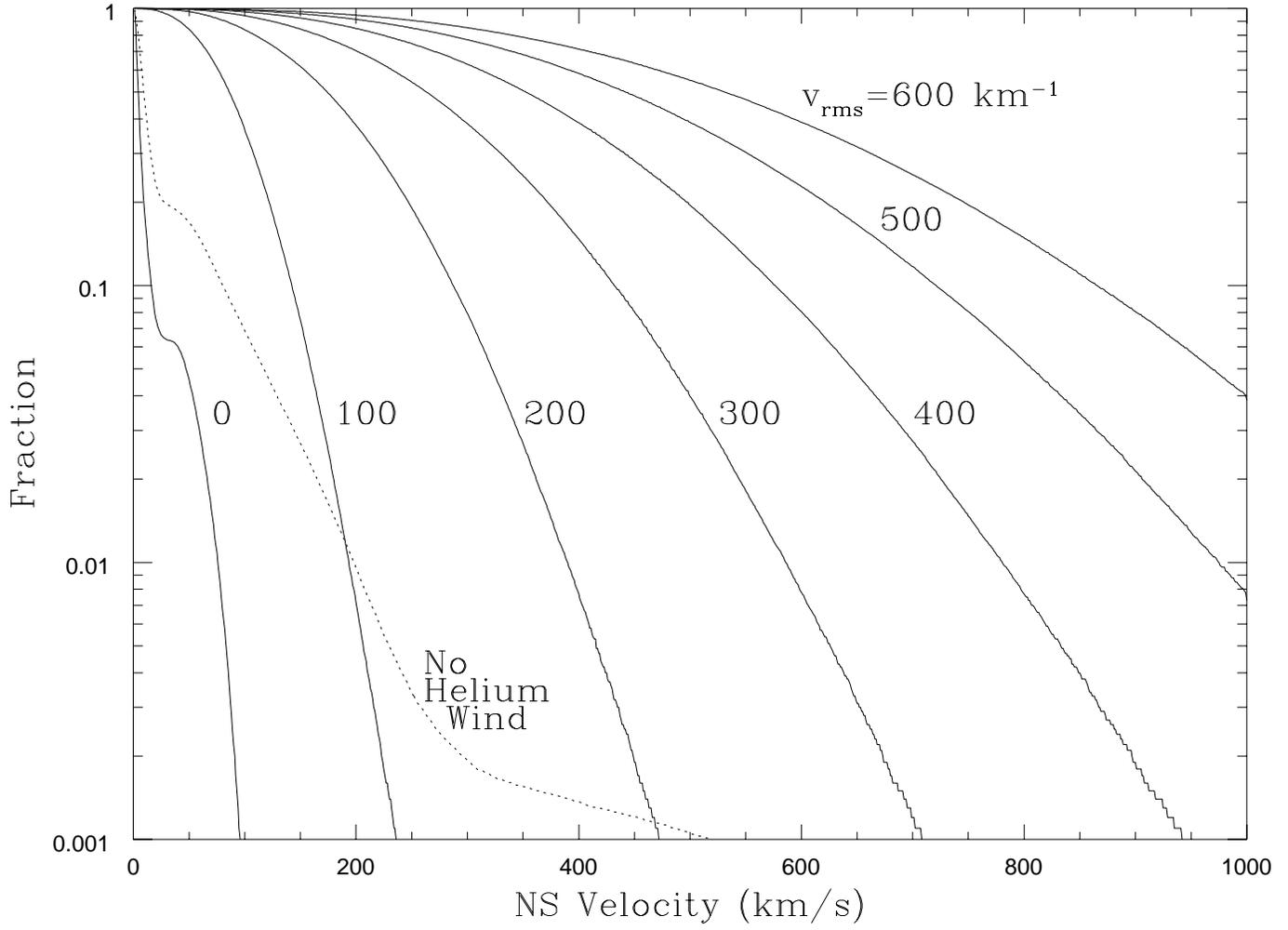}{7in}{-90}{70}{70}{-280}{520}                              
\caption{Fraction of Neutron Stars with a velocity greater than a given velocity 
versus that velocity for the series of Maxwellian kick distributions.  The dotted 
line that shows a zero 
kick simulation with no mass loss from He winds is included to demonstrate how 
binary effects can change the velocity distribution.}
\end{figure}

\begin{figure}
\plotfiddle{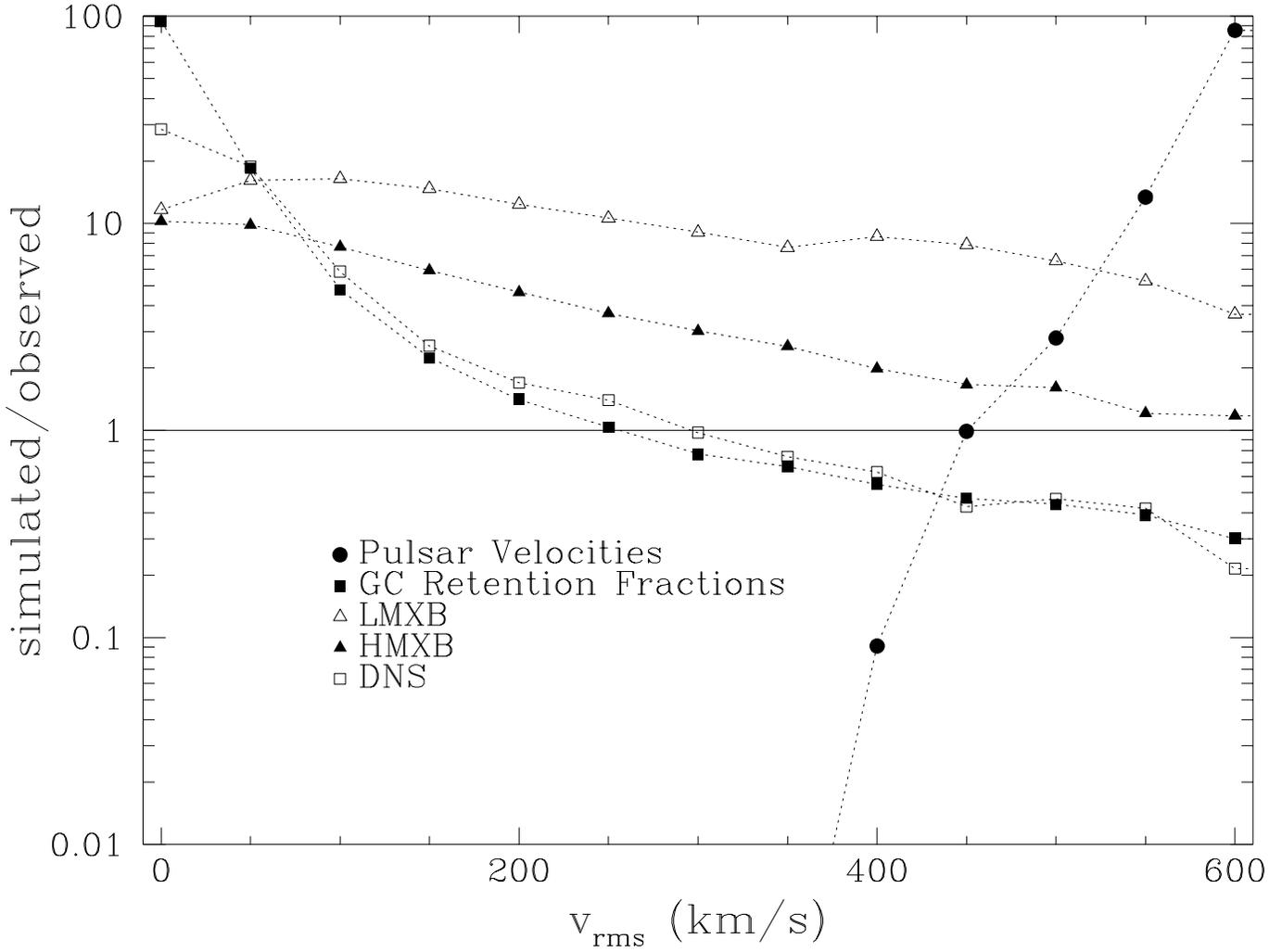}{7in}{-90}{70}{70}{-280}{520}
\caption{The same as Fig. 8, but with $q_0=0.15$ and $M_{l,u}$=10,100$M_{\odot}$.}
\end{figure}

\begin{figure}
\plotfiddle{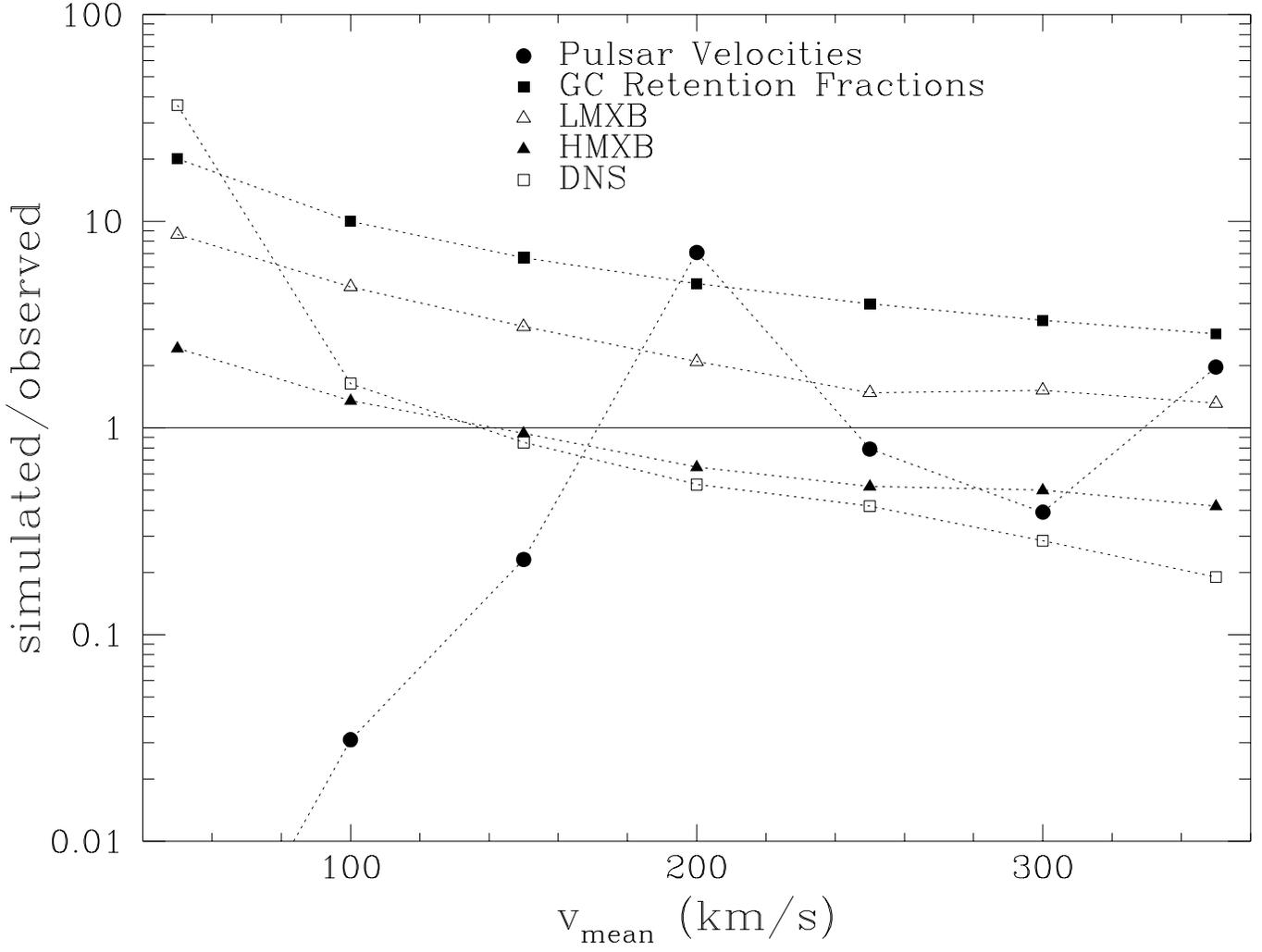}{7in}{-90}{70}{70}{-280}{520}
\caption{The same as Fig. 8, but for flat kick distributions with $q_0=0.15$.}
\end{figure}

\begin{figure}
\plotfiddle{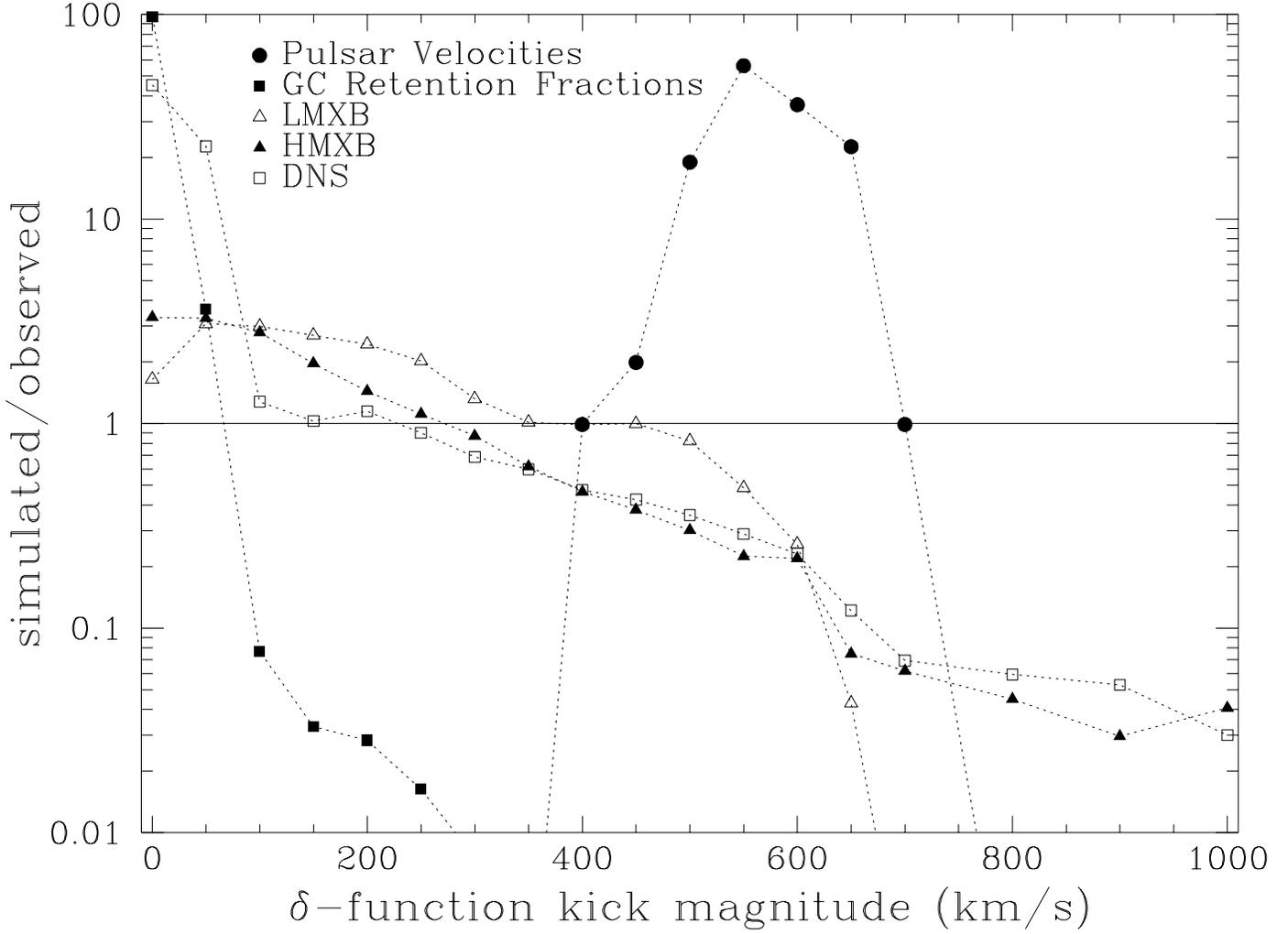}{7in}{-90}{70}{70}{-280}{520}
\caption{The same as Fig. 8, but for $\delta$-function kicks.}
\end{figure}

\begin{figure}
\plotfiddle{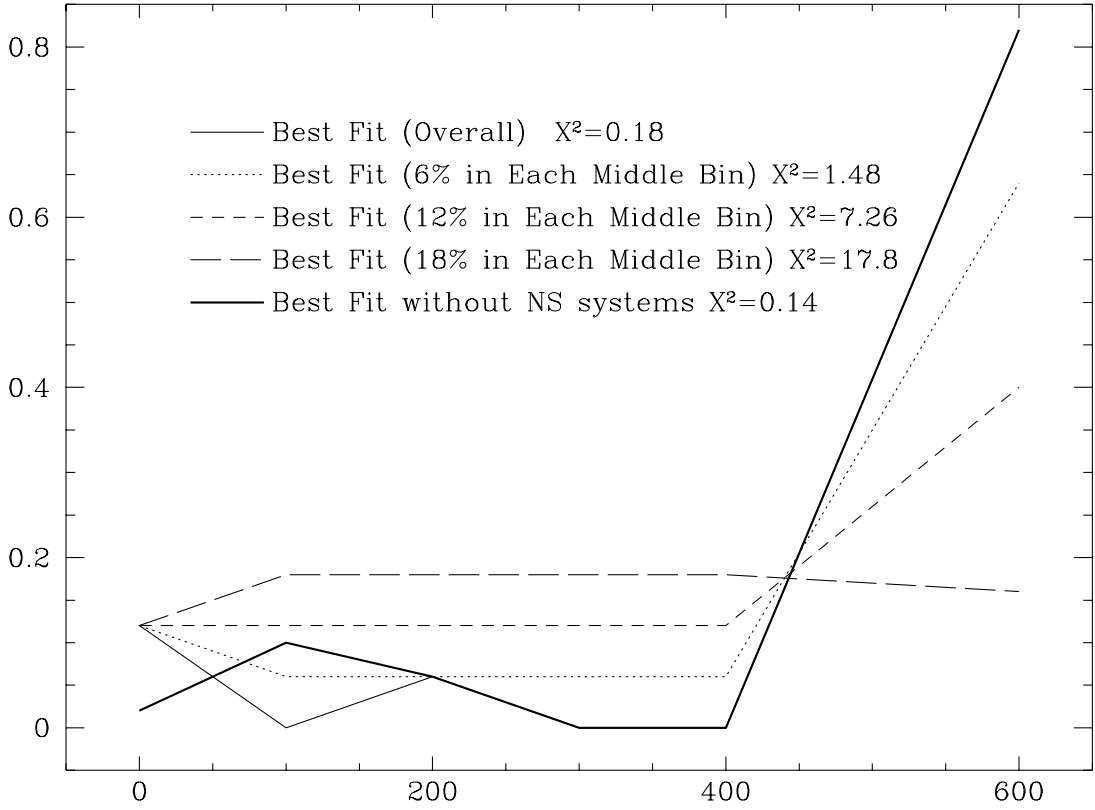}{7in}{-90}{60}{60}{-250}{450}
\caption{Kick distributions with fraction of neutron stars versus velocity bin.  
The solid line denotes the best-fitting overall kick profile.  The remaining 
curves are constrained by requiring that the middle range of velocities be 
non-zero to varying degrees.}
\end{figure}

\begin{figure}
\plotfiddle{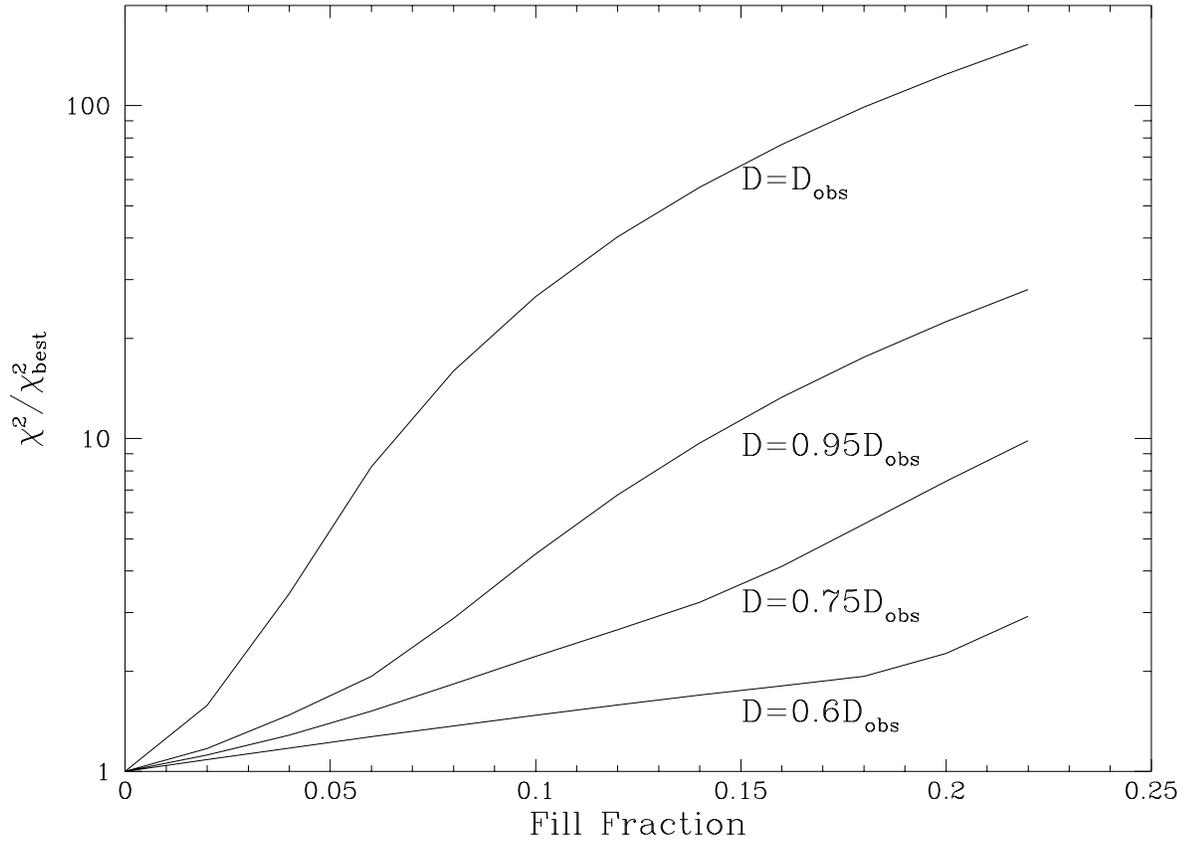}{7in}{-90}{60}{60}{-250}{450}
\caption{$\chi^2$ residuals versus the degree to which the middle velocities are 
filled, normalized by the residuals from the best-fitting overall distribution.  
The curves represent a range of scale factors for the distance.}
\end{figure}

\begin{figure}
\plotfiddle{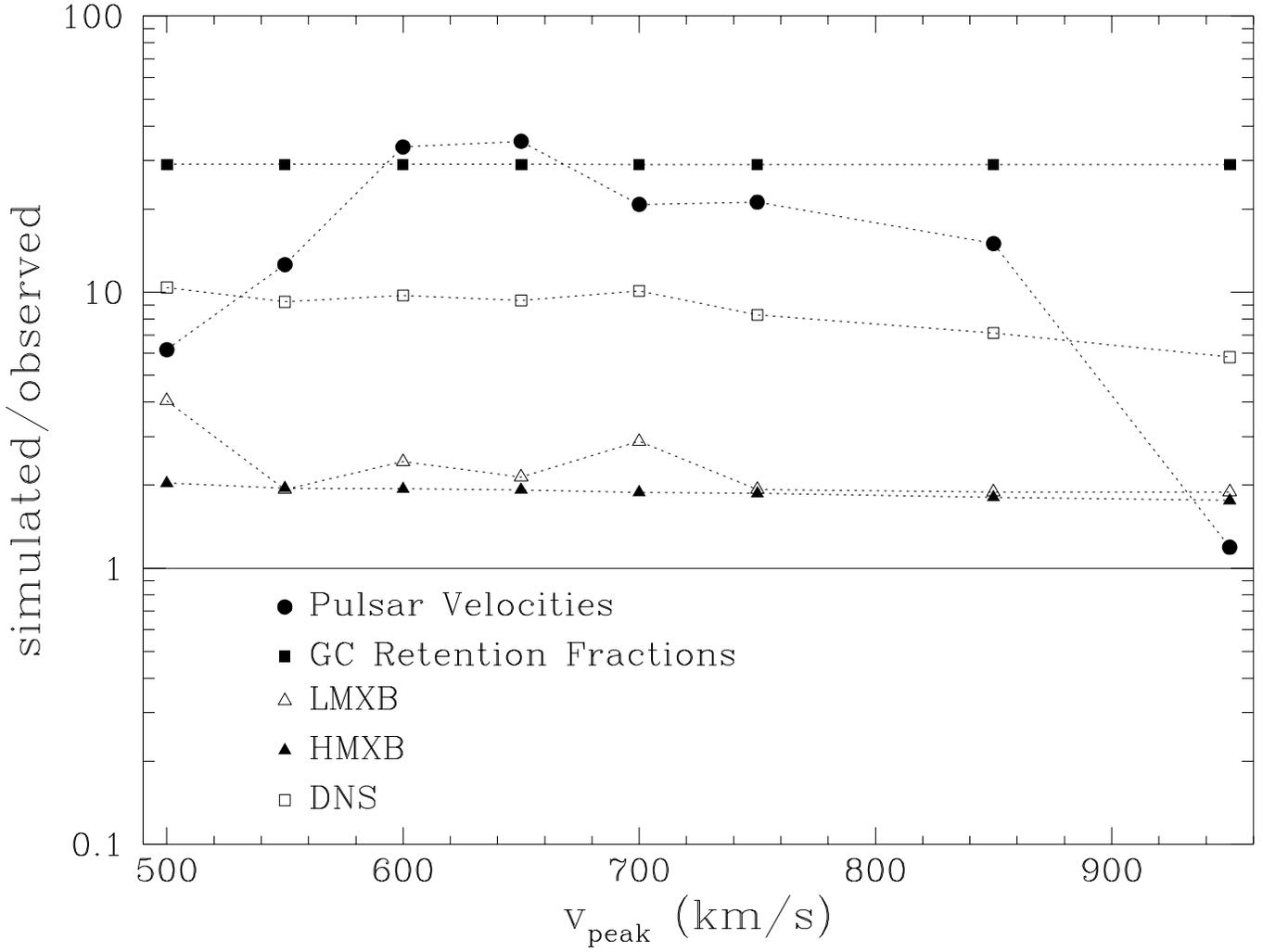}{7in}{-90}{70}{70}{-280}{520}
\caption{Same as Fig 8. with a double-peaked $\delta$-function 
distribution:  30\% at 0 km/s and 70\% at the value on the 
plot.  Again, we use $q_0=0.15$.}
\end{figure}

\begin{figure}
\plotfiddle{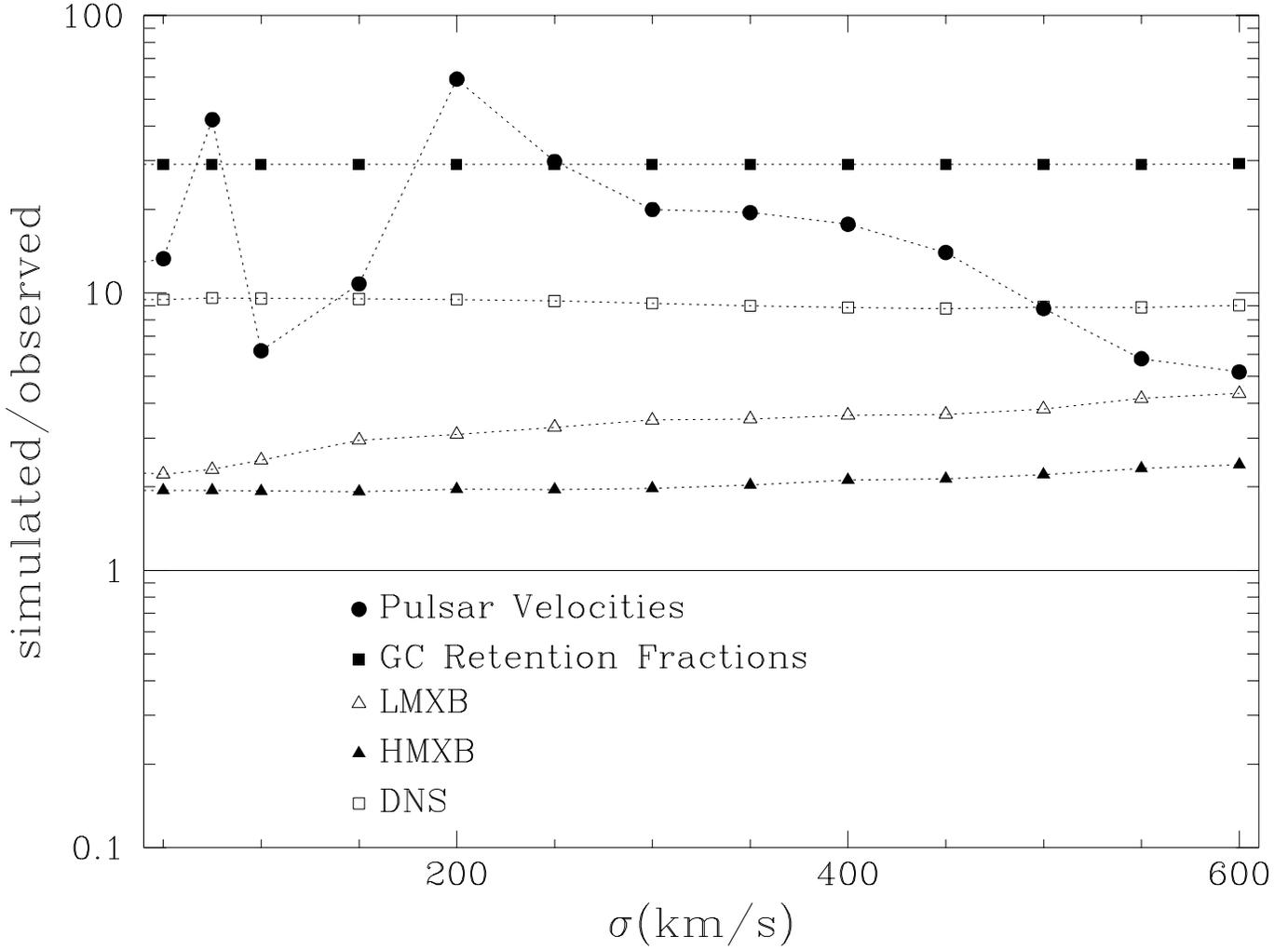}{7in}{-90}{70}{70}{-280}{520}
\caption{Same as Fig 8. with a double-peaked distribution:  
30\% at 0 km/s and 70\% with a flat distribution 
with a mean at 625 km/s and a dispersion:  $\sigma$.  
Again, we use $q_0=0.15$.}
\end{figure}

\begin{figure}
\plotfiddle{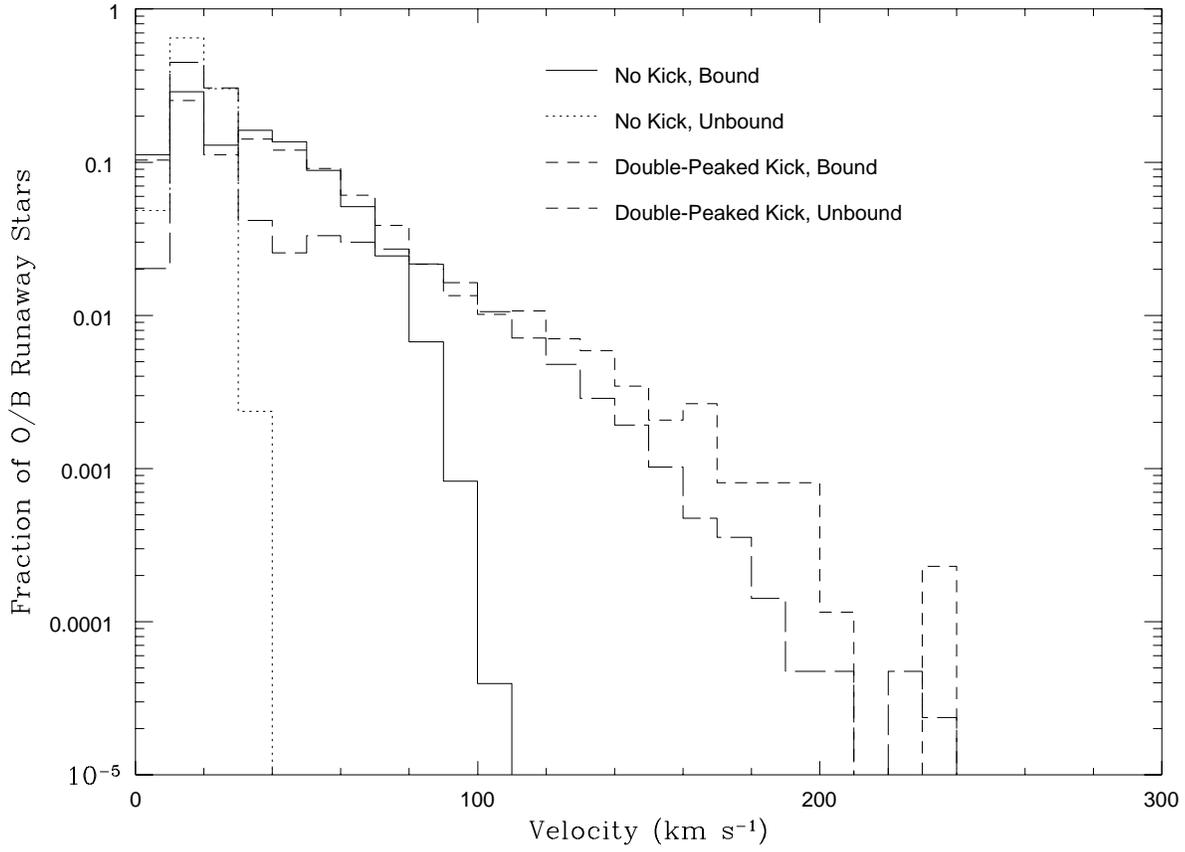}{6in}{-90}{60}{60}{-260}{450}
\caption{The velocity distribution of O/B stars for simulations with 
no kick (solid line - bound systems, dotted line - unbound O/B stars) 
and for our best fitting double peaked distribution (long-dashed line - 
bound systems, short-dashed line - unbound systems.)}
\end{figure}

\end{document}